\documentclass[usenatbib,usegraphicx,useAMS]{mn2e}

\def\lsim{\mathrel{\hbox{\rlap{\hbox{\lower4pt\hbox{$\sim$}}}\hbox{$<$}}}}
\def\gsim{\mathrel{\hbox{\rlap{\hbox{\lower4pt\hbox{$\sim$}}}\hbox{$>$}}}}

\title[Galaxy morphology in the $\Lambda$CDM cosmology]{Galaxy morphology in the $\Lambda$CDM cosmology}

\author[O.~H.~Parry et al.]{O.~H.~Parry\thanks{E-mail:o.h.parry@durham.ac.uk}
V.~R.~Eke and C.~S.~Frenk\\ 
Institute for Computational Cosmology, Department of Physics,
University of Durham, Science Laboratories, South Road, Durham DH1
3LE} 

\def\apj{ApJ}
\def\mnras{MNRAS}
\def\pasp{PASP}
\def\nat{Nature}
\def\apjs{ApJS}
\def\apjl{ApJL}
\def\aj{AJ}

\def\araa{ARA\&A}
\def\aap{A\&A}
\def\etal{{\rm et al}}

\begin{document}

\date{Accepted 200? . Received 200? ; in original form 200?}

\pagerange{\pageref{firstpage}--\pageref{lastpage}}
\pubyear{2008}

\maketitle
\label{firstpage}

\begin{abstract}
We investigate the origins of galaxy morphology (defined by bulge-to-total K-band luminosity) in the $\Lambda$CDM cosmology using two galaxy formation models based on the Millennium simulation, one by Bower et al. (the Durham model) and the other by De Lucia \& Blaizot (the MPA model).  Both models have had considerable success in reproducing a number of observed properties of the local and high redshift universe, including star formation rates, the stellar mass function and the luminosity function out to $z\sim5$.  There are many similarities, but also fundamental disagreements in the predictions of the two models for galaxy morphology.  For example, taking into account uncertainties  in the available observational data, both produce a realistic morphological mix today, but its evolution is very different.  A main cause of this and other differences is the treatment of disk instabilities which play a more prominent role in the Durham model.  Our analysis confirms previous theoretical predictions that elliptical galaxies form most of their stars before the bulk of the galaxy is assembled.  Spirals tend to have later `assembly' times than ellipticals as a consequence of in-situ star formation.  With the exception of the brightest ellipticals (stellar mass $M_{*}\gsim 2.5\times 10^{11}h^{-1}M_{\odot}$), we find that major mergers are {\it not} the primary mechanism by which most spheroids (ellipticals and spiral bulges) assemble their mass.  In fact, the majority of ellipticals (and the overwhelming majority of spirals) never experience a major merger (above the resolution limit of our simulation.) Most ellipticals and spiral bulges acquire their stellar mass through minor mergers or disk instabilities.  These conclusions are common to both the MPA and Durham models.  The rotation properties of spheroids may help to constrain the importance of disk instabilities in these models.

\end{abstract}

\begin{keywords}
galaxies: formation -- galaxies: evolution -- galaxies: bulges.
\end{keywords} 

\section{Introduction} \label{sec:intro}
The morphological type of a galaxy describes more than merely its physical appearance; it reveals the key processes that have shaped it and continue to affect its evolution.  Hubble's well established classification scheme for galaxies~\citep{Hubble1926,Hubble1936} has survived with only modest modification~\citep{DV1959,VDB1960,Sandage1961}, precisely because the disks and spheroids on which it relies are fundamental products of the formation process.  Understanding the origins of such structures is therefore of central importance if we are to build up a coherent picture of galaxy formation and evolution.
Modern observational facilities like the Hubble Space Telescope have made it possible to probe progressively higher redshifts and, together with large surveys like the Sloan Digital Sky Survey (SDSS), are beginning to reveal how galaxy morphology changes with time.  A significant trend in the low redshift universe is the `density-morphology' relation~\citep{Dressler1980,Postman1984,Dressler1997}, which shows that denser regions contain proportionally more early-type galaxies.  This relation holds at least back to $z\sim1$~\citep{Postman2005}.  At higher redshifts, between $z\sim0.5$ and 2, there appears to be a rapid breakdown of the Hubble classification system~\citep{VanDenBergh2002,Papovich2005} with the proportion of galaxies classed as irregulars or peculiars rising steadily, particularly at faint magnitudes~\citep[e.g.,][]{Brinchmann2000}.  Nonetheless, a substantial fraction of giant spirals are observed at moderate redshifts~\citep[e.g.,][]{Ravindranath2004}.  Some authors have claimed that this population shows very little evolution up to the present, implying that at least some Hubble sequence galaxies were in place at $z\sim1$~\citep{Marleau1998,Ravindranath2004}, whilst others counter that many of the higher redshift sample would have become bulge dominated in the intervening time, so they are not directly comparable to their $z=0$ counterparts~\citep{Bell2008}.  Before $z=2$, the majority of observed galaxies are either highly irregular or centrally compact~\citep{Giavalisco1996,Dickinson2000,Daddi2004}.  The difficulty in linking populations seen at different epochs has ensured that a clear evolutionary picture for morphology remains elusive.

At present, direct simulation of the processes that determine morphology remains too computationally taxing to perform for large samples of galaxies (although significant progress is being made in this respect; see for example \citealp{Croft2008}).  Semi-analytic techniques constitute a convenient tool to capture the most important characteristics of these simulations, whilst also providing the statistics required to perform detailed analyses.  The galaxy formation models considered in this paper are described in $\S$~\ref{sec:MSMODELS}.

Our current understanding of galaxy formation tells us a little about how morphology develops.  It is well known, for instance, that angular momentum acquired by a proto-galactic gas cloud through tidal torques operating during its expansion and collapse will naturally cause it to settle into a disk~\citep{Hoyle1949,Peebles1969,White1984}.  Beyond this though, the combination of mechanisms required to explain morphological change are poorly understood.  \citet{Toomre1977} was the first to put forward the idea that mergers could be responsible for the formation of elliptical galaxies.  Indeed, the importance of mergers in the hierarchical cold dark matter (CDM) model makes them a natural candidate for such a role~\citep{Frenk1985}.  Early CDM galaxy formation models stressed the possibility that, through a combination of mergers and gas accretion, a single galaxy could, at various stages in its lifetime, be identified with several morphological types on the Hubble sequence~\citep{White1991}, a view supported by gas-dynamical simulations~\citet{Steinmetz2002}.  Numerous simulations have confirmed at least that major mergers, or very strong gravitational encounters, can lead to the complete destruction of a disk and the formation of a spheroid with properties similar to elliptical galaxies~\citep[e.g.,][]{White1978,Negroponte83,Barnes1992,Barnes1996,Bekki1997,Naab1999,Springel2005b,Naab2003,Cox2006}. Even in virialised clusters where the relative velocities of galaxies make strong gravitational interactions of any kind infrequent, mergers will still have affected groups collapsing to form the cluster, including those it accretes at later times. 

Other phenomena act to change morphology through their effects on gas content and star formation.  For instance,~\emph{strangulation} (also called~\emph{suffocation}) occurs when a galaxy with a gaseous atmosphere falls into a larger structure, causing gas cooling to be heavily supressed~\citep{Larson1980,White1991}.  In the formation models of \citet{Bower2006} and \citet{DeLuciaBlaizot2007} considered here, ~\emph{strangulation} is modelled by stripping all infalling (satellite) galaxies of their hot gas supplies, preventing further cooling.  In combination with efficient feedback from supernovae, this can lead to the reddening of a satellite's stellar population on a relatively short timescale.  Two further effects that are not included in these two models are \emph{ram pressure stripping}~\citep{GunnGott1972,Abadi1999} and \emph{harassment}~\citep{Moore1996}, both of which are most important for cluster galaxies.  The former involves the stripping of gas as galaxies move through the intracluster medium (ICM), while the latter results from a series of intense, high speed gravitational encounters that can cause bursts of star formation and lead to disk disruption.  Recent additions to the Durham galaxy formation model analysed in this paper \citep{Bower2006} include a detailed prescription for ram pressure stripping \citep{Font2008}.
In this paper, we use an online database to examine two different galaxy formation models implemented on the $\Lambda$CDM Millennium Simulation, the first by \citet{DeLuciaBlaizot2007} (hereafter the MPA model) and the second by \citet{Bower2006} (hereafter the Durham model).  We look at the morphological content of each model as a function of redshift and track the evolution of the luminosity function of each type.  We go on to determine how a galaxy's formation epoch, and that of its host dark matter halo, affect its final morphology and investigate the role played by major mergers, minor mergers and disk instabilities.  Our main goal is to investigate galaxies of all morphological types and to compare and contrast the predictions of the two models, but we also carry out limited comparisons to observations.  The rest of the paper is laid out as follows.  In $\S$\ref{sec:MSMODELS}, we describe the N-body simulation and semi-analytic models that form the basis of this analysis. $\S$\ref{sec:DATA} discusses the methods we use to select and retrieve the data and compares some basic properties of morphology in the models with SDSS data.  Finally, our results and conclusions are presented in $\S$\ref{sec:RESULTS} and $\S$\ref{sec:CONCLUSIONS}.

\section{The N-Body simulation and semi-analytic models} \label{sec:MSMODELS}
\subsection{The Millennium simulation} \label{sec:MS}

Both galaxy formation models investigated in this paper are based on the Millennium Simulation (MS) of the evolution of cold dark matter in a representative cosmological volume.  Completed by the Virgo Consortium in 2004, the MS remains the largest published N-body cosmological simulation, representing a significant step forward in terms of spatial, temporal and mass resolution (see~\citet{Springel2005sup} for a detailed description).  It uses $2160^{3}$ ($\sim10^{10}$) particles to model a cubic region of space,$500h^{-1}$ Mpc on a side,\footnote{Here, h is defined by $H_0=h\times100\rm{kms}^{-1}\rm{Mpc}^{-1}$.} and tracks the evolution of structure between $z=127$ and the present day.  The initial  configuration of the particles was constructed using a $\Lambda$CDM power spectrum consistent with cosmological parameters from 2dFGRS~\citep{Colless2001} and first year WMAP~\citep{Spergel2003} data.  The parameters used were $\Omega_m$ = 0.25, $\Omega_b$ = 0.045, $h = 0.73$, $\Omega_{\Lambda}$ = 0.75, $n = 1$ and $\sigma_8$ = 0.9\footnote{Here, $\Omega_m$, $\Omega_b$ and $\Omega_{\Lambda}$ are the densities of all matter, baryons and dark energy respectively, in units of the critical density ($\rho_{crit}=3H^{2}/8\pi G$), $n$ is the initial power spectrum slope and $\sigma_8$ is the rms overdensity predicted today by linear theory for a sphere of radius $8h^{-1}$Mpc.}  at $z=0$.  Particle positions and velocities were recorded in 64 output snapshots.

A friends-of-friends (FOF) algorithm~\citep{Davis1985} was used `on the fly' to identify groups, linking adjacent particles separated by less than 0.2 times the mean interparticle separation.  Groups with 20 or more particles (corresponding to a minimum halo mass of $1.7\times 10^{10}h^{-1}M_{\odot}$) were retained for further analysis.  The FOF groups were further processed using a version of the SUBFIND algorithm~\citep{Springel2001} that identifies gravitationally bound locally overdense regions, which to which we refer here as subhaloes. In the MPA model, a halo, by definition, includes all subhaloes within a FOF group.  In the Durham model, on the other hand, a FOF group may be split into more than one halo if: i) a subhalo is outside twice the half-mass radius of the original halo, or ii) a subhalo was identified in a previous snapshot as part of a different halo and has retained $75\%$ or more of its mass~\citep{Harker2006}.  Both of these conditions are designed to prevent two haloes that are temporarily joined by a tenuous particle bridge being linked together.  The latter condition is applied on the basis that a subhalo loses significant mass from its outer layers when falling into a more massive counterpart, but substantially less if it is merely a close encounter. In both models, the descendant of a subhalo is identified by following its most tightly bound particles and the descendant of a halo is that which contains most of its most massive subhalo.  With the merger trees of these bound structures thus defined, the two models employ different techniques to populate them with galaxies.
For a detailed description of the two semi-analytic models analysed here, the reader is referred to the papers in which they were originally presented \citep{Bower2006,DeLuciaBlaizot2007} and references therein, particularly \citet{Croton2006} and \citet{Cole2000}.  The following subsections summarise aspects  that are particularly relevant to the discussion of morphology.

\subsection{Gas cooling, star formation and supernova feedback} \label{sec:COOL}

The treatment of radiative gas cooling in both models essentially follows the framework set out by \citet{White1991}.  The amount of gas initially available to cool is a fixed fraction of the mass of the dark matter halo\footnote{The reionisation of the gas at early times modifies this fraction somewhat for lower mass haloes in both models.}. This gas shock-heats as it falls into the halo's potential well, attaining the virial temperature.  As it cools and collapses,  the angular momentum that it acquired prior to turnaround causes it naturally to settle into a disk, which cools fastest in the centre where the density is highest.  For low mass haloes, cooling is very rapid and the supply of cold gas is limited purely by the rate at which gas can free-fall onto the disk.  For high mass haloes, a \emph{quasi-static} hot atmosphere forms, from which gas can be accreted as it cools in the dense central regions.

An important difference between the two models is in how they calculate the instantaneous rate of gas cooling.  \citet{Bower2006} allow gas to accrete onto the cold disk if it lies within a `cooling radius', that at which the local cooling time is equal to the `age' of the halo.  They define the halo's age as the time since it last doubled in mass \citep{Cole2000}. \citet{DeLuciaBlaizot2007}, on the other hand, calculate a rate based on the amount of gas that has a cooling time less than the halo dynamical time, which defines an alternative cooling radius.  The Durham  prescription typically results in higher cold gas masses than the MPA  prescription by a factor of $\sim2$ at $z\lsim5$ in central galaxies,  rising to a factor of $\sim3.5$  by $z\sim0$.  Stars are assumed to form from the cold disk gas, with Kennicutt \citep{Kennicutt1983} and Chabrier \citep{Chabrier2003} initial mass functions in the Durham and MPA cases respectively.  In both models, the stellar population synthesis model of \citet{Bruzual2003} is used to obtain stellar population properties.  Star formation in the disk affects its development significantly.  Stellar winds and supernovae increase the metallicity of the disk material from which further stars form and inject substantial amounts of energy into their surroundings. This can drive gas and metals out of the disk, changing the composition of the surrounding hot gas and altering its cooling time.

\subsection{Mergers and disk instabilities} \label{sec:MERGE}

Two mechanisms incorporated into the Durham and MPA models are responsible for disrupting the stellar disks that would otherwise form as described above -- galaxy mergers and disk instabilities.  Galaxy mergers occur as a direct consequence of, but are distinct from, the mergers of the haloes that host them.  As described in $\S$~\ref{sec:MS}, a halo that falls into a more massive system may survive for some time thereafter, as a gravitationally bound subhalo. Hence, the less massive (satellite) galaxy can be followed explicitly until tidal effects disrupt its subhalo sufficiently for it to drop below the 20 particle resolution limit.  From this point onwards, in the MPA model, the galaxy is associated with the most bound particle of the subhalo just before it became unresolved.  The satellite's orbit is assumed to decay through dynamical friction against the halo material until it merges with the more massive (central or primary) galaxy.  Merger timescales are determined differently in the Durham model, with the satellite's orbit chosen at random from the cosmological distribution derived by \citet{Benson2005}, as soon as the halo merger has taken place and a dynamical friction timescale calculated accordingly.

In the Durham model, the result of a merger is dictated by the relative masses of the merging galaxies and the gas content of the primary.  `Major' mergers ($M_{sat}/M_{pri}\ge0.3$), completely disrupt any existing disk, producing a spheroidal remnant that contains all the stars from its progenitor galaxies.  Any gas present forms stars in a burst, which are also added to the new spheroid, with some fraction of gas being returned to the hot halo through supernovae feedback.  Minor mergers ($M_{sat}/M_{pri}<$0.3) leave the structure of the primary intact, simply adding the stars from the satellite to the bulge of the primary and its gas to the disk, but may also initiate a burst if $M_{sat}/M_{pri}\ge$0.1 and the primary has sufficient gas in its disk ($M_{gas}/M_{disk}>$0.1).  The fraction of the available gas that is turned into stars is dictated by star formation and feedback rules, applied over the timescale of the burst.

The MPA model assumes starbursts to occur in all mergers.  The mass fraction of gas available to take part is a power-law function of the satellite-central mass ratio $(m_{sat}/m_{central})$.  Other than this, the mechanics of mergers are largely similar to the Durham model.  Major mergers are defined in the same fashion and also result in the formation of a spheroid; minor mergers add stars to the primary's bulge and gas to its disk.

In galaxies with strongly self-gravitating disks, that is, where the mass of the disk itself dominates the gravitational potential, the system is unstable to perturbations from nearby satellites or dark matter clumps~\citep{Efstathiou1982,Mo1998}.  Both models regard a disk to have become unstable when
\begin{equation}
V_{max}/(G M_{disk}/r_{disk})^{1/2}\leq1, 
\label{eqn:Inst}
\end{equation}
where $M_{disk}$ and $r_{disk}$ are the disk's total mass and radius.  In the original formulation by \citet{Efstathiou1982}, $V_{max}$ was the maximum of the rotation curve, but this is approximated by the halo virial velocity in the MPA model and the disk's velocity at its half mass radius in the Durham model.  Although the prescriptions will typically agree on whether a particular disk is unstable or not, they deal with the situation somewhat differently.  In the MPA model, there is effectively a partial collapse, intended to model the formation of a bar.  Mass is moved from the disk into the spheroid until the stability of the system, defined by eqn.~\ref{eqn:Inst} is re-established.  In the Durham model, instabilities result in the complete collapse of the disk into a spheroid, the size of which is dictated by the rotational energy of the disk just prior to collapse. As with a major merger, a starburst is induced, such that the resulting spheroid contains the original disk's stars, plus those formed in the burst.  This more catastrophic outcome is assumed to result from orbital resonances and stellar scattering in the barred system causing it to collapse entirely.

\subsection{Black holes and AGN feedback} \label{sec:BH}

Prescriptions for supermassive black hole (BH) growth and feedback from active galactic nuclei (AGN) are relatively recent additions to both models, motivated by the growing body of evidence linking properties of a galaxy's bulge to the mass of its central BH~\citep[e.g.,][]{Magorrian1998,Ferrarese2000,Kormendy2001,Marconi2003,Haring2006}. Energy produced through accretion of material onto the BH is assumed to restrict gas cooling, particularly in higher mass haloes, impacting on galaxy luminosity, colour and morphology.  Different implementations of AGN feedback could, for instance, dictate whether or not an elliptical galaxy is able to regrow a stellar disk after a major merger.

The prescription used in the Durham model for AGN feedback and BH growth is described in full in \citet{Malbon2007}.  In the Durham model, BH growth is assumed to take place through four main channels. i) \emph{Major merger-driven accretion} - Tidal forces act to drive disk gas into central regions, fuelling the BH.  A constant fraction of the accreted gas, $f_{BH}$, (tuned to match the amplitude of the $M_{BH}-M_{Bulge}$ relation) is added to the BH. ii) \emph{Instability-driven accretion} - A fraction, $f_{BH}$, of the gas in the collapsing disk is added to the BH. iii) \emph{BH-BH mergers} - When two galaxies merge, their BHs are assumed to merge.  Mass loss due to the radiation of gravitational waves is neglected and the mass of the new BH is the sum of the progenitor masses plus any gas accreted. iv) \emph{Accretion from cooling flows} - In sufficiently massive haloes, where \emph{quasi-hydrostatic cooling} is taking place, the BH can accrete mass from the cooling flow.  This last mechanism is closely associated with the implementation of AGN feedback in the model.  In haloes where a quasi-static hot halo has formed, the energy generated through accretion is assumed to couple efficiently with the hot halo gas.  If the energy output of the BH (some constant fraction of its Eddington luminosity) exceeds the rate at which the gas can radiate away energy, then no further gas is allowed to cool.  Hence, as soon as a large cooling flow builds up, the feedback becomes sufficient to cut it off.  This behaviour plays a key role in reproducing the observed K and B-band luminosity functions out to $z\sim5$.

The MPA implementation also assumes that gas inflows associated with mergers lead to increased BH accretion rates.  Consistent with their starburst model, such inflows are presumed to occur in all mergers, with the accreted fraction of the total gas dependent on the satellite-central mass ratio.  A simple phenomenological model is used to describe accretion from quasi-hydrostatic cooling flows; the mass accretion rate ($\dot{m}$) depends on the mass of the black hole, the hot gas fraction and the virial velocity of the halo.  The luminosity of the BH at any given time is simply $\eta\dot{m}c^2$, where $\eta\simeq 0.1$ is the efficiency of mass-energy conversion typically expected close to the event-horizon.  Based on this energy output, an adjusted cooling rate is calculated, such that it declines smoothly as the amount of mass accreted rises.  As in the Durham model, BH-BH mergers occur and are modelled in a similar fashion.  However, no additional accretion is considered during a disk collapse.
\section{The data} \label{sec:DATA}
\subsection{The Millennium Run database}

All of the data analysed here comes from the online Millennium Run database\footnote{Hosted at http://www.g-vo.org/MyMillennium, with a mirror site at http://galaxy-catalogue.dur.ac.uk:8080/MyMillennium (Both require registration for full access).}, developed by the German Astrophysical Virtual Observatory (GAVO), which can be accessed through the use of structured query language (SQL)~\citep{Lemson2006} and contains galaxy and halo data from the Durham and MPA models.  Various properties are associated with each halo, for instance, mass, redshift, position, velocity, number of subhaloes, etc.  Galaxies additionally carry many directly and indirectly observable properties, such as observer and rest frame magnitudes in several passbands, stellar mass and radial extent of morphological components.  The database has been constructed to make the retrieval of an object's merger tree as efficient as possible (see~\citet{Lemson2006b} for a detailed description).  This has been achieved by assigning unique IDs in a `depth-first' manner, such that the progenitors of a given object have IDs lying between that of the object itself and an index it carries called the \emph{lastprogenitorid}.  Additional indices identify each object's immediate descendant and largest progenitor.  An additional index in the Durham tables identifies every galaxy's main branch.

\subsection{Definition of morphology and construction of galaxy samples}

To describe a galaxy's morphology we use the bulge-total (B-T) ratio of absolute, K-band, rest frame luminosity ($R=L_{K,bulge}/L_{K,total}$).  Although most photometric measurements are made in the B-band, we chose the K-band because it reflects the stellar mass quite closely, even at moderate redshifts \citep[e.g.][]{Kauffmann1998,Lacey08}. This choice has the further advantage that the predictions are then relatively insensitive to uncertain details of the current star formation and reddening corrections in the model. Galaxies are divided into three broad morphological types: ``spirals" ($R<0.4$), ``S0s'' ($0.4\leq R\leq0.6$) and ``ellipticals" ($R>0.6$). This classification is, to some extent, arbitrary, but at least in the B-band \citet{Tran2001} found that galaxies classified as late Hubble types in the NASA Extragalactic Database are well by a rest-frame B-band B-T ratio of less than 0.4, consistent with our definition of spirals.

In addition, we split our two $z=0$ galaxy populations by luminosity, with the division at $M_{K}-\rm{5logh}=-22.17$, i.e. one magnitude fainter than the characteristic luminosity in the K-band \citep[e.g.]{Cole01,Smith2008}.  Throughout this work, we refer to these as the \emph{bright} and \emph{faint} populations.  Where the r-band is also shown, we divide the populations at $M_{r}-\rm{5logh}=-19.83$~\citep{Blanton2001} and define morphological classes according to the same B-T ratios as in theK-band.  ~\citet{DeLuciaBlaizot2007} note that, in order to assign a morphology to a galaxy with confidence, its merger history must be well resolved.  They determine that this condition imposes a lower limit of $4\times10^{9}h^{-1}M_{\odot}$ in stellar mass, which we apply consistently to the bright and faint populations in both models.  This cut has virtually no impact on the bright population but it reduces the numbers in the faint population by $\sim92-94\%$, leaving us with 1,298,118 bright and 1,889,131 faint galaxies in the MPA model and 1,280,154 bright, 1,598,908 faint galaxies in the Durham model.  

\section{Overview of morphological characteristics}
\subsection{Evolution of the morphological mix} \label{sec:morph_frac}
\begin{figure*}
$\begin{array}{c@{\hspace{-0cm}}c@{\hspace{-0cm}}}
\multicolumn{1}{l}{\mbox{\bf}} & 
	\multicolumn{1}{l}{\mbox{\bf}} \\
	\includegraphics[width=0.39\textwidth]{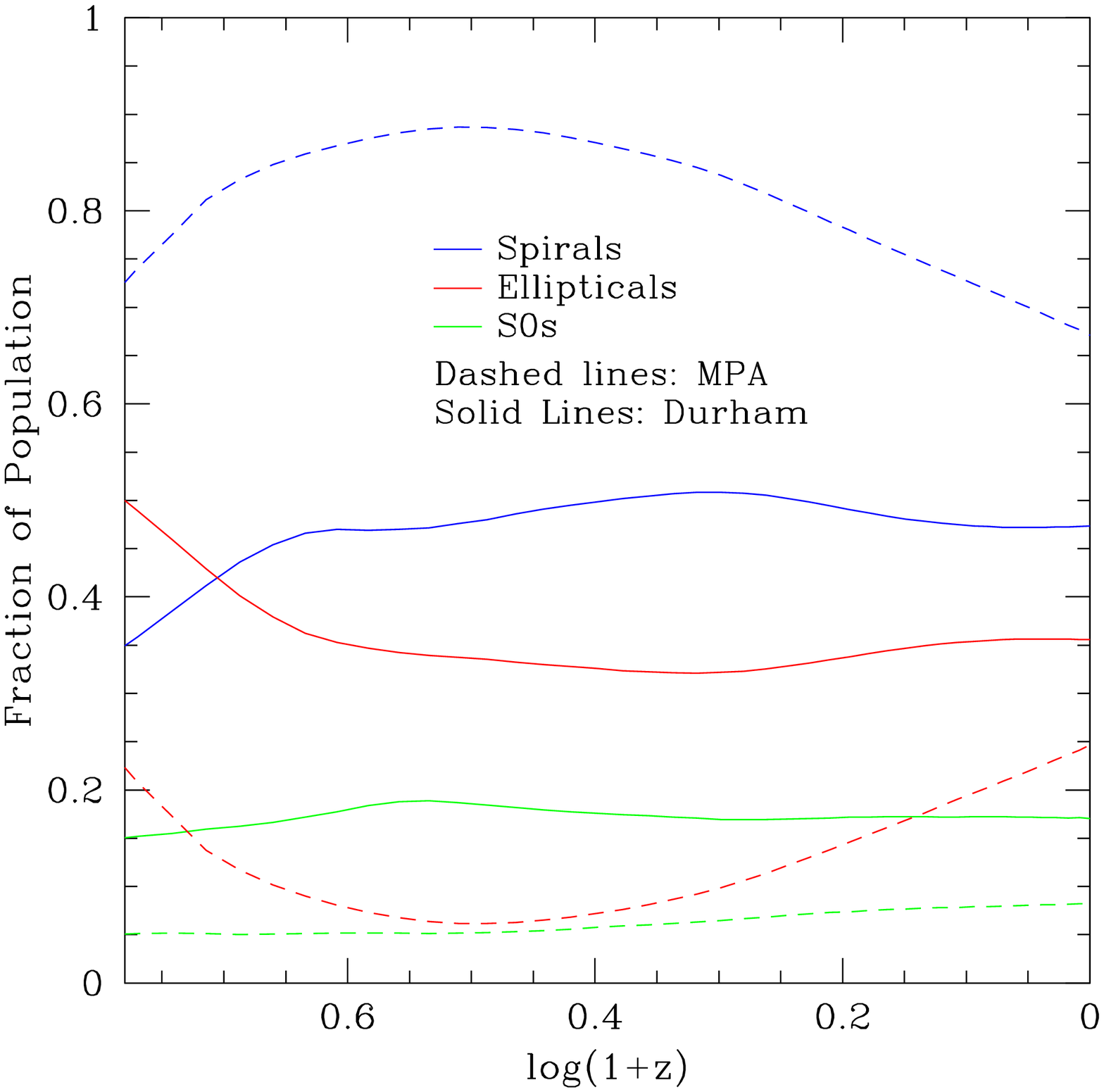} &
	\includegraphics[width=0.39\textwidth]{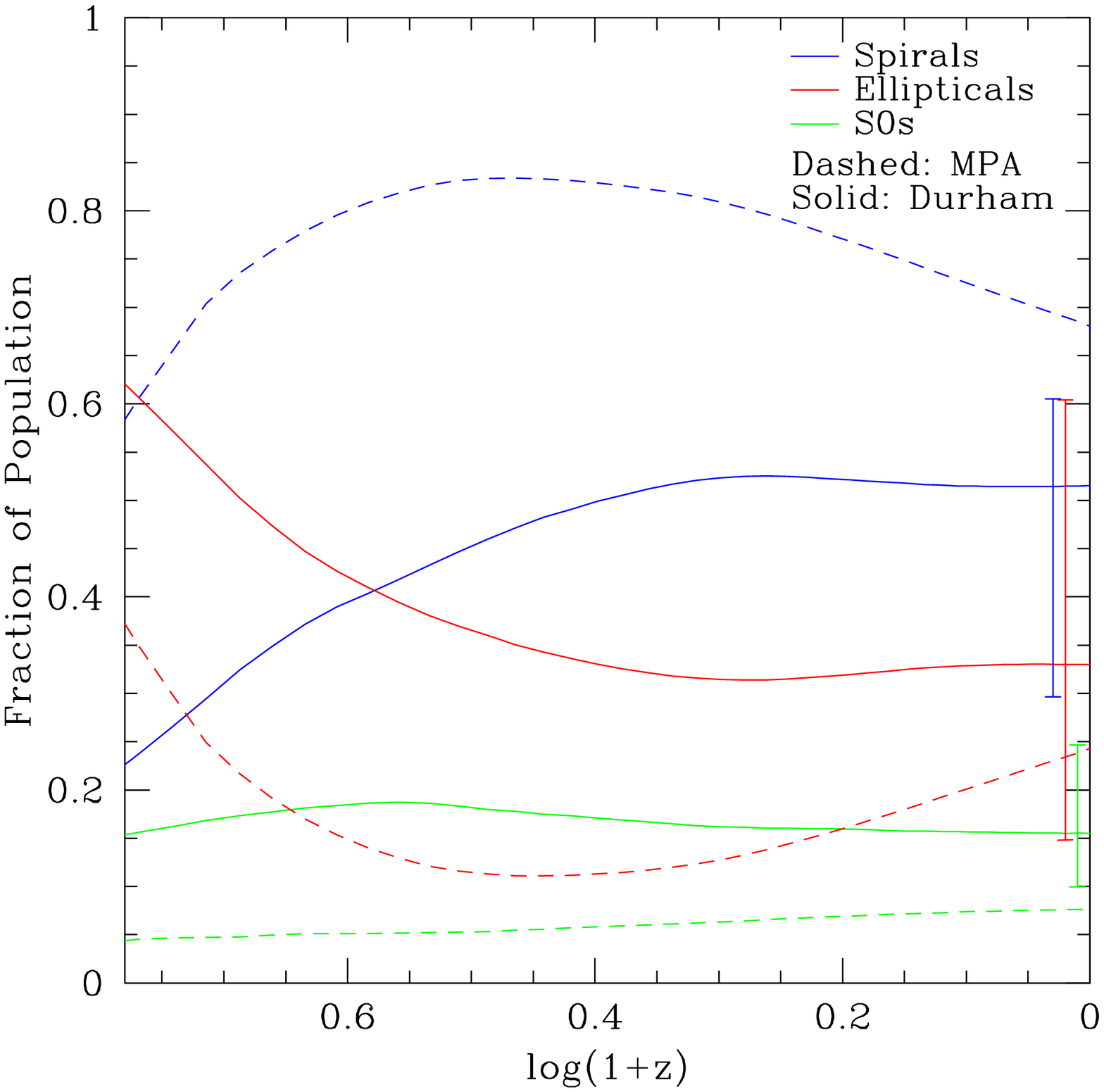}\\
\end{array}$
\caption{The fraction of bright galaxies classified as spirals (blue), S0s (green) and ellipticals (red) in the Durham (solid lines) and MPA (dashed lines) models as a function of redshift.  In the left panel, morphology is defined using rest-frame K-band light and in the right panel using rest-frame r-band light.  The bars on the right of the r-band panel represent the possible range in each fraction at $z=0$, calculated using the data of \citet{Benson2007}.  We explain the origin of these observed ranges in our discussion of Fig.~\ref{fig:BT}.  Disks are the dominant structures in the MPA model at all times, whereas the Durham model produces a transition from a bulge-dominated phase before $z\sim3-4$ to a disk-dominated phase therefafter.  This strikingly different behaviour appears to arise from the contrasting treatment of disk instabilities in the two models.  Despite the differences, the two prescriptions arrive at fairly similar fractions at $z=0$.}
\label{fig:morph_frac}
\end{figure*}

With our galaxy samples thus defined, we now consider some basic properties of the two models.  Firstly, we examine their morphological content, that is, the relative fractions of each morphology at a given redshift.  Fig.~\ref{fig:morph_frac} tracks the fraction of the total bright galaxy populations in the K-band (left panel) and r-band (right panel) that are spirals, S0s and ellipticals in each model.  The r-band is included here in order to compare with the observational data of \citet{Benson2007}. 

\begin{figure*}
$\begin{array}{c@{\hspace{-0cm}}c@{\hspace{-0cm}}c@{\hspace{-0cm}}}
\multicolumn{1}{l}{\mbox{\bf}} &
	\multicolumn{1}{l}{\mbox{\bf}} \\
	\includegraphics[width=0.32\textwidth]{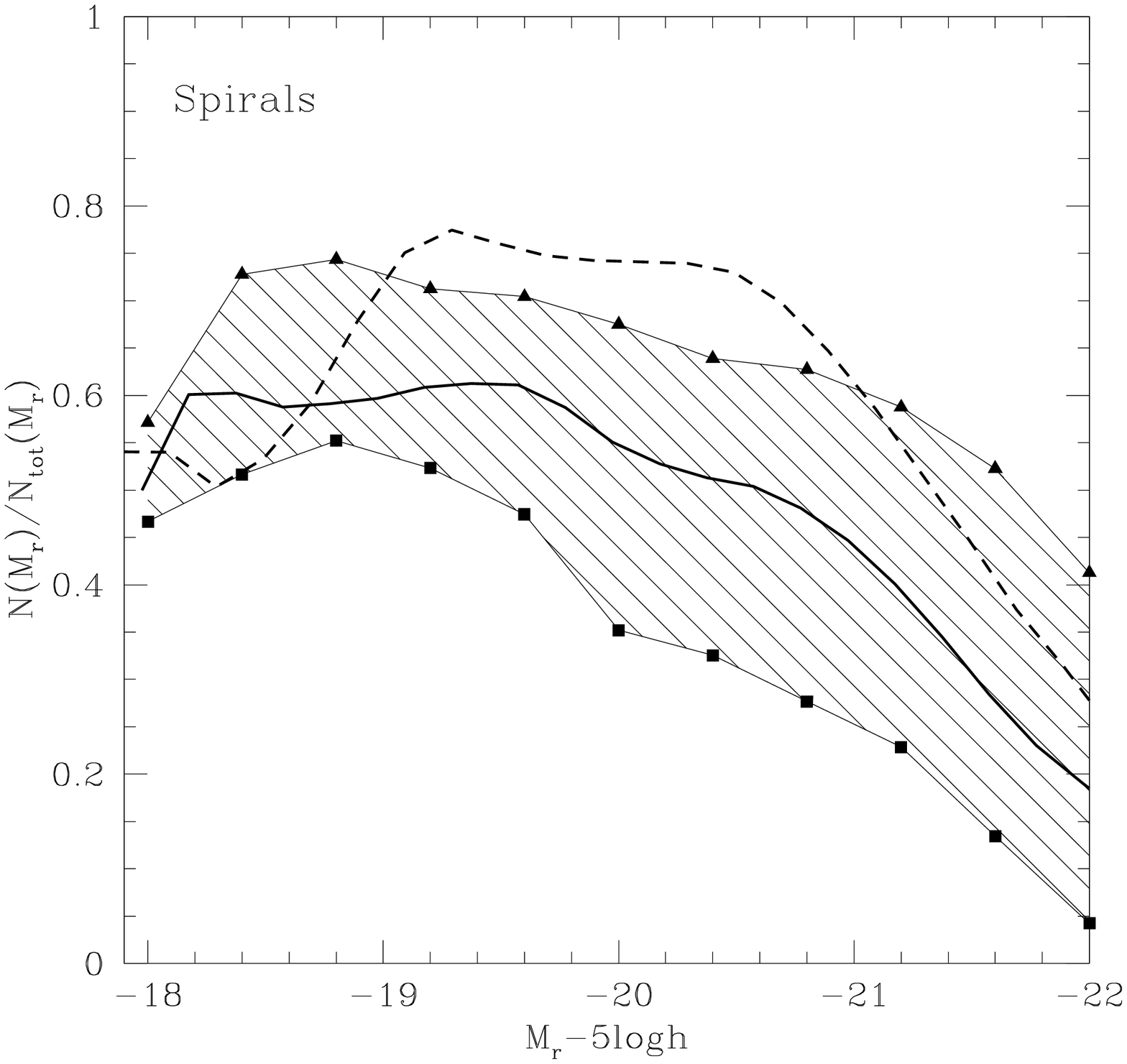} &
	\includegraphics[width=0.32\textwidth]{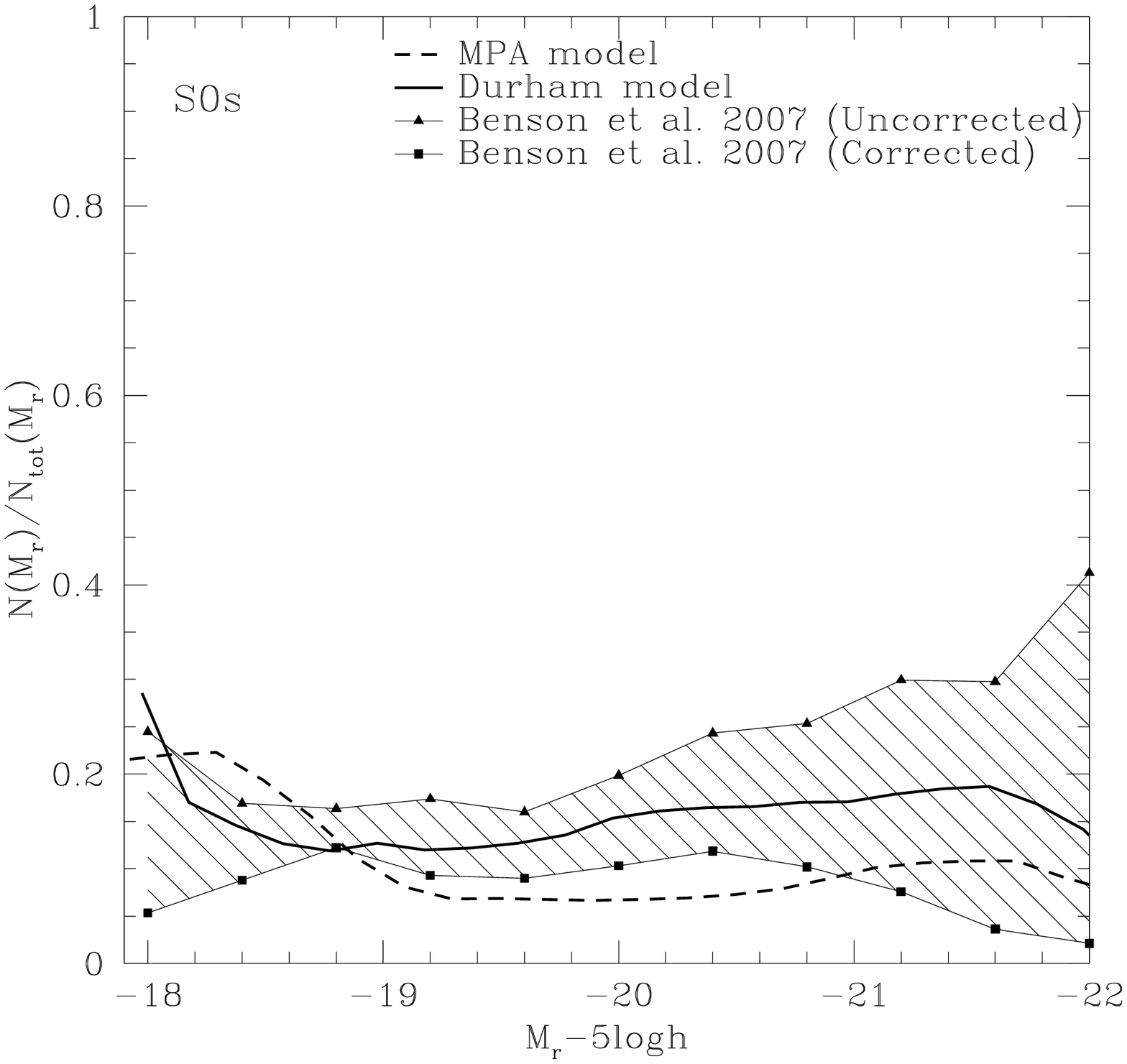} &
	\includegraphics[width=0.32\textwidth]{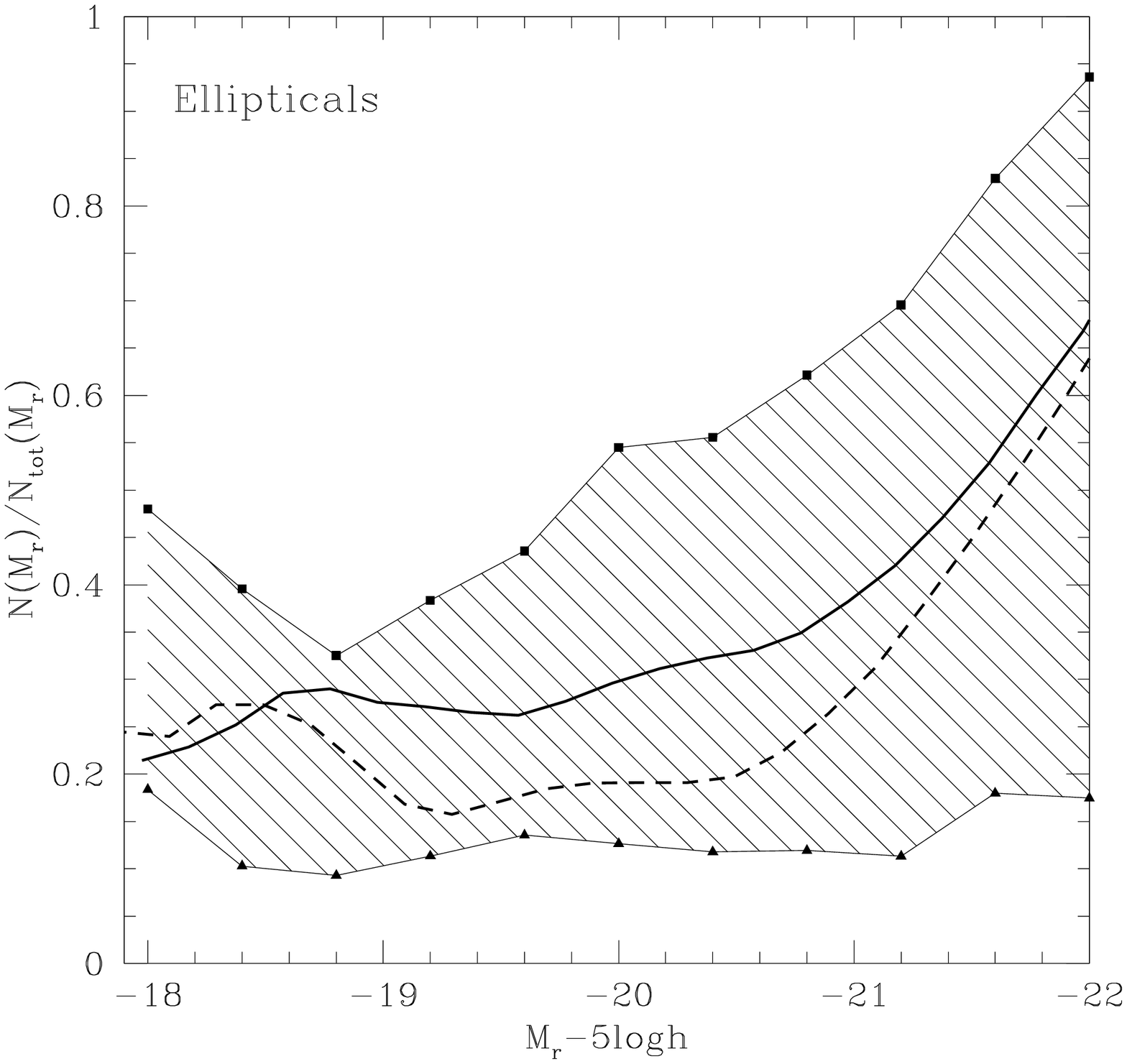} \\
	\mbox{\bf (a)} & \mbox{\bf (b)} & \mbox{\bf (c)}\\
\end{array}$
\caption{The fraction of galaxies at $z=0$ that are spirals {\bf(a)}, S0s {\bf(b)} and ellipticals {\bf(c)}, as a function of rest-frame r-band magnitude in the MPA model (thick dashed line),  Durham model (thick solid line) and in the SDSS data~\citep{Benson2007}(connected points).  The two sets of points represent the data before (triangles) and after (squares) a  correction for biases introduced in the fitting process. This correction is explained in the text. The upper and lower magnitude limits of the plot are dictated by the data.  Both models reproduce the general trend and typically lie between the uncorrected and corrected data, although the MPA model appears to have too many spirals and not enough S0s over a range of magnitudes.} 
\label{fig:BT} 
\end{figure*}

Clearly, in both bands, the two models differ considerably, particularly at high redshifts.  The Durham model predicts a substantial population of ellipticals at early times, with steady evolution to a disk-dominated phase after $z\sim3-4$.  In contrast, the MPA model shows more modest evolution, with spirals remaining the most prevalent structures throughout.  Even so, the two prescriptions result in fairly similar present day morphological compositions (see table \ref{tab:morph_frac} for K-band fractions).  Fractions in the Durham model are consistent with the ranges defined by the observational data, but the MPA model seems to produce too many spirals and not enough S0s.  We will explain the origin of these ranges in our discussion of Fig.~\ref{fig:BT}.  The difference between bands for both models is a small offset toward higher B/T values in the r-band, which is a consequence of an increased sensitivity to dust obscuration compared to the K-band, which acts to dim disk light. 

Considering instead the faint galaxy population ($M_{K}-\rm{5logh}>-22.17$, not shown here), we find a very similar result.  The plot looks almost identical for MPA galaxies and similar for Durham galaxies, though there is an offset of about $10-15\%$ in favour of more spirals at the expense of S0s and ellipticals.     

\begin{table}
 	\centering
	\caption{Percentage contribution of each morphology to the present day populations of the MPA and Durham models.} 
	\label{tab:morph_frac}
	\resizebox{8.7cm}{!} {
	\begin{tabular}{c cc cc}  
		\hline
		& \multicolumn{2}{c}{Durham} & \multicolumn{2}{c}{MPA}\\
		\hline
		& Bright & Faint & Bright & Faint\\
		& $(M_{K}-\rm{5logh}<-22.17)$ & $(M_{K}-\rm{5logh}>-22.17)$ & $(M_{K}-\rm{5logh}<-22.17)$ & $(M_{K}-\rm{5logh}>-22.17)$\\
		\hline
		Spirals&51&62&67&72 \\
		S0&15&12&8&11 \\
		Elliptical&34&26&25&17 \\
		\hline
	\end{tabular}
	}
\end{table}

The disagreement at high redshift evident in Fig.~\ref{fig:morph_frac} may be explained by the contrasting treatment of disk instabilities in the two models.  As outlined in $\S$~\ref{sec:MERGE}, instabilities trigger a total collapse of the galactic disk in the Durham model, but only a partial ``buckling" in the MPA model.  As some of the results in $\S$~\ref{sec:RESULTS} demonstrate, this distinction appears to be the root cause of several differences in galaxy morphology in the two models. 

Several attempts have been made to quantify morphology through bulge-total measurements at low redshift \citep[e.g.,][]{Tasca2005,Benson2007,Driver2007,Gadotti2008}.  In particular, \citet{Benson2007} showed that the Durham model \citep{Bower2006} reproduces the luminosity function of disks and spheroids\footnote{Spheroids were taken to include both the bulges of spirals and elliptical galaxies.} remarkably accurately. Fig.~\ref{fig:BT} uses their SDSS data to compare the fraction of spirals, S0s and ellipticals in the local universe ($z<0.12$) as a function of (AB, rest-frame) r-band magnitude, with fractions from the two models at $z=0$.  For this plot, morphologies in the models are defined using rest-frame r-band light for consistency with the data.

Benson et al. determined B/T ratios for their sample of $\sim9000$ SDSS galaxies by fitting two component light profiles (bulge + disk) to each image.  Analysis of their initial dataset showed that the distribution of disk inclination angles appeared to be biased, leading them to conclude that their code was fitting face-on disks where the bulge varied significantly from the anticipated De Vaucouleurs profile.  They applied a statistical correction which attempted to make the distribution of inclination angles uniform, setting B/T to 1 for those galaxies left over after the correction.  As Benson et al. point out, this correction will likely overestimate the prevalence of ellipticals and underestimate the number of spirals, but the magnitude of the uncertainty is very difficult to quantify.  We have therefore chosen to show the data before and after the applied correction to illustrate the range of uncertainty for each fraction.  The ranges at $z=0$ in Fig. \ref{fig:morph_frac} are obtained by integrating the corrected and uncorrected fractions for each type, multiplied by the r-band luminosity function \citep{Blanton2001}.

{With this in mind, we see in Fig.~\ref{fig:BT} that both models reproduce the general trend in the data - ellipticals are more prevalent at brighter magnitudes, and typically fall between the corrected and uncorrected datasets.  The fraction of spirals in the MPA model exceeds the uncorrected data at intermediate magnitudes and the fraction of S0s falls below the corrected data over the same range, but by an amount less than size of the Poisson errors on each datasets (not shown).  Consequently, neither model is obviously discrepant with the data, although clearly smaller uncertainties would be desirable in order to place tighter constraints.


\subsection{The morphology-dependent luminosity function}

The luminosity function provides an important census of any galaxy population.  Fig. \ref{fig:morphlf0} shows the rest-frame B and K-band luminosity functions for each morphological type at $z=0$, the solid lines representing the Durham model and the dashed lines the MPA model.  A galaxy's K-band luminosity can be used as a reasonable proxy for its stellar mass~\citep{Kauffmann1998,Brinchmann2000,Bell2001}, whilst the B-band is biased toward younger stellar populations and hence is indicative of recent star formation.  On both plots, the black vertical dashed line marks the magnitude faintward of which the imposed mass cut at $4\times10^{9}h^{-1}M_{\odot}$ depletes galaxy densities by $>5\%$. 

\begin{figure*}
$\begin{array}{c@{\hspace{-0cm}}c@{\hspace{-0cm}}}
\multicolumn{1}{l}{\mbox{\bf}} & 
	\multicolumn{1}{l}{\mbox{\bf}} \\
	\includegraphics[width=0.39\textwidth]{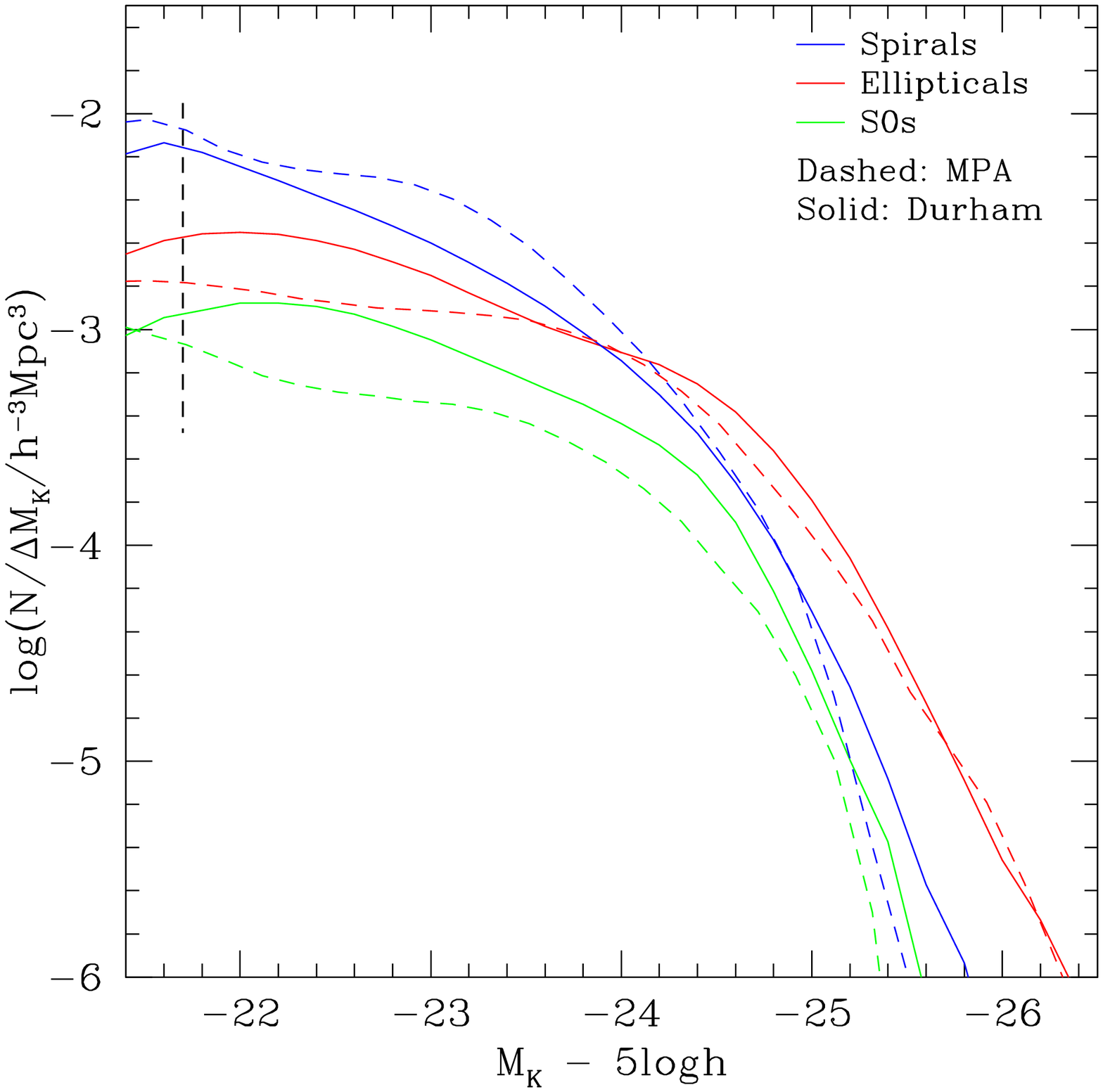} &
	\includegraphics[width=0.39\textwidth]{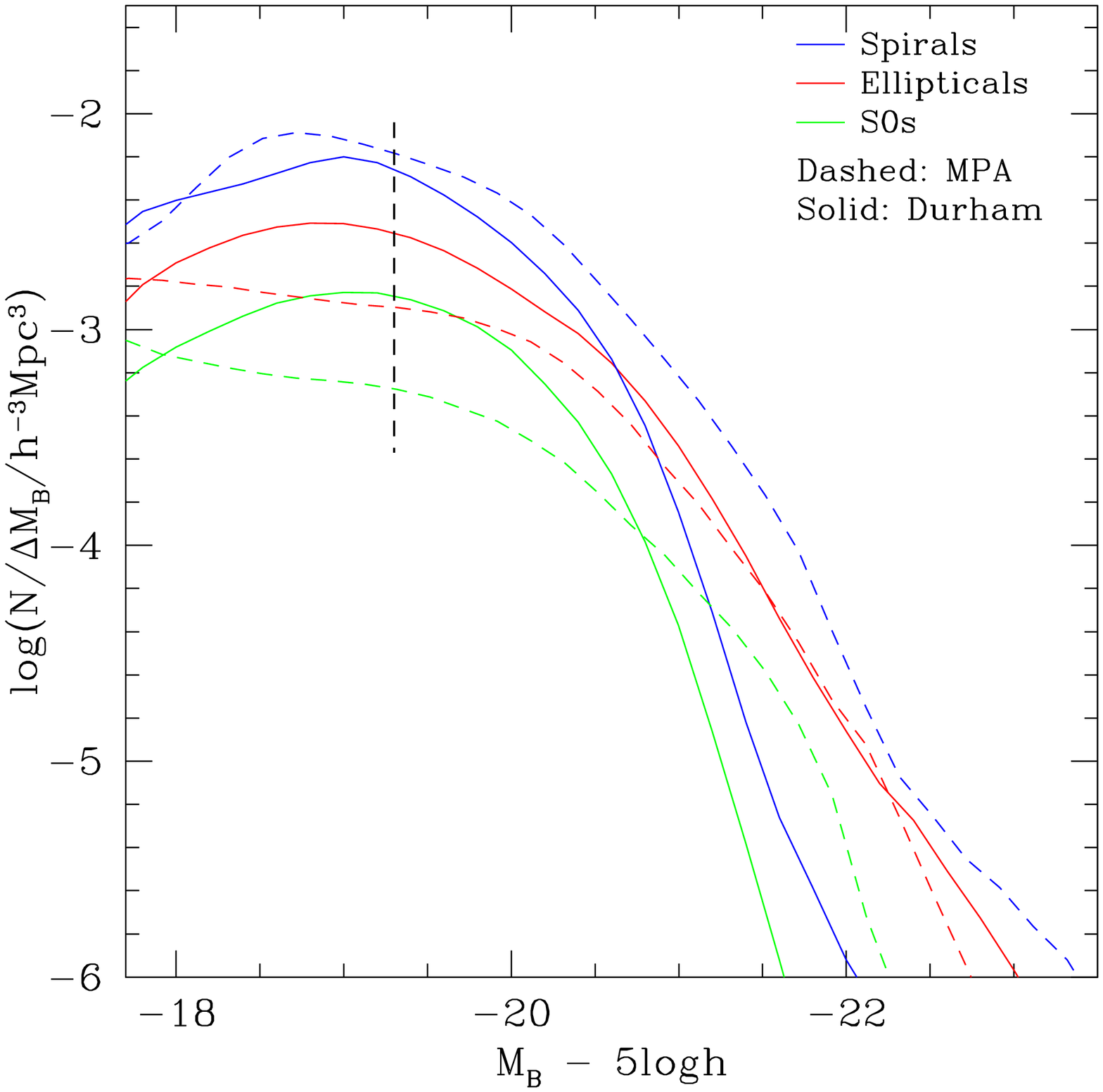}\\
\end{array}$
\caption{The $z=0$ rest-frame luminosity functions in the K-band (left panel) and B-band (right panel) for each morphology.  The solid lines show the Durham model and the dashed lines the MPA model.  The black dashed lines indicate the magnitude faintwards of which the imposed mass cut reduces galaxy number densities by $>5\%$.}  
\label{fig:morphlf0}
\end{figure*} 

Both models produce luminosity functions which, qualitatively at least, share some features with comparable observational data.  For instance, the most massive (brightest K-band) galaxies are ellipticals and spirals have a somewhat fainter characteristic magnitude~\citep{Nakamura2003}.  We also note that spiral galaxies in the MPA model are typically brighter in the B-band than those in the Durham model, suggesting that they have significantly more ongoing star formation or less dust obscuration.  In both models, ellipticals are the most common morphology at high ($M<M_{*}$) K-band luminosities, but at the bright end of the B-band LF, the MPA model has at least an order of magnitude more more spirals than the Durham model.

Fig.~\ref{fig:morphlf} shows the redshift evolution of the K and B-band luminosity functions for each morphology as ratios to the $z=0$ functions in Fig. \ref{fig:morphlf0}.  The reader should note that the bright-end extent of these curves is limited by that of the $z=0$ data and the faint-end is cut off where the imposed mass cut reduces number densities by $>5\%$.  In the K-band, there is very little evolution in the Durham model between $z\simeq 2$ and $z\simeq 0.5$ for all spirals and S0s and for all but the brightest ellipticals, suggesting that a substantial proportion of their mass was already in place at that time.  At the bright end, there is already a sizeable population of ellipticals at $z=5$ in both models.  The early dominance of ellipticals in the Durham model seen in Fig. \ref{fig:morph_frac} comes mostly from galaxies excluded from this plot by the faint-end cut off.  The number density of bright ellipticals in the Durham model declines slightly from $z\simeq 2$ to $z\simeq 0.5$, a result both of mergers between ellipticals and their transformation into S0s or spirals for those able to grow a disk.  The trend is in the opposite sense for the MPA model, indicating that the extra spirals already in place at $z=5$ compared to the Durham model make merging a process that typically increases elliptical number density.  Evolution of the faint-end K-band is more significant in the MPA model, particularly for ellipticals.  This may reflect weaker supernovae feedback allowing more star formation in satellite galaxies.

The Durham B-band functions are consistent with a peak in star formation somewhere around $z=2$, for spirals and S0s, and  somewhat earlier for ellipticals.  In the MPA model, there is a steady  decline in B-band light for all types since $z\simeq 5$, suggestive of consistently  earlier star formation than in the Durham model.  The history of star  formation in the different types is examined more closely in the next section.

\begin{figure*} 
$\begin{array}{c@{\hspace{-0cm}}c@{\hspace{-0cm}}}
\multicolumn{1}{l}{\mbox{\bf}} &
	\multicolumn{1}{l}{\mbox{\bf}} \\
\mbox{\bf\large K-band} & \mbox{\bf\large B-band}\\
	\includegraphics[width=0.38\textwidth]{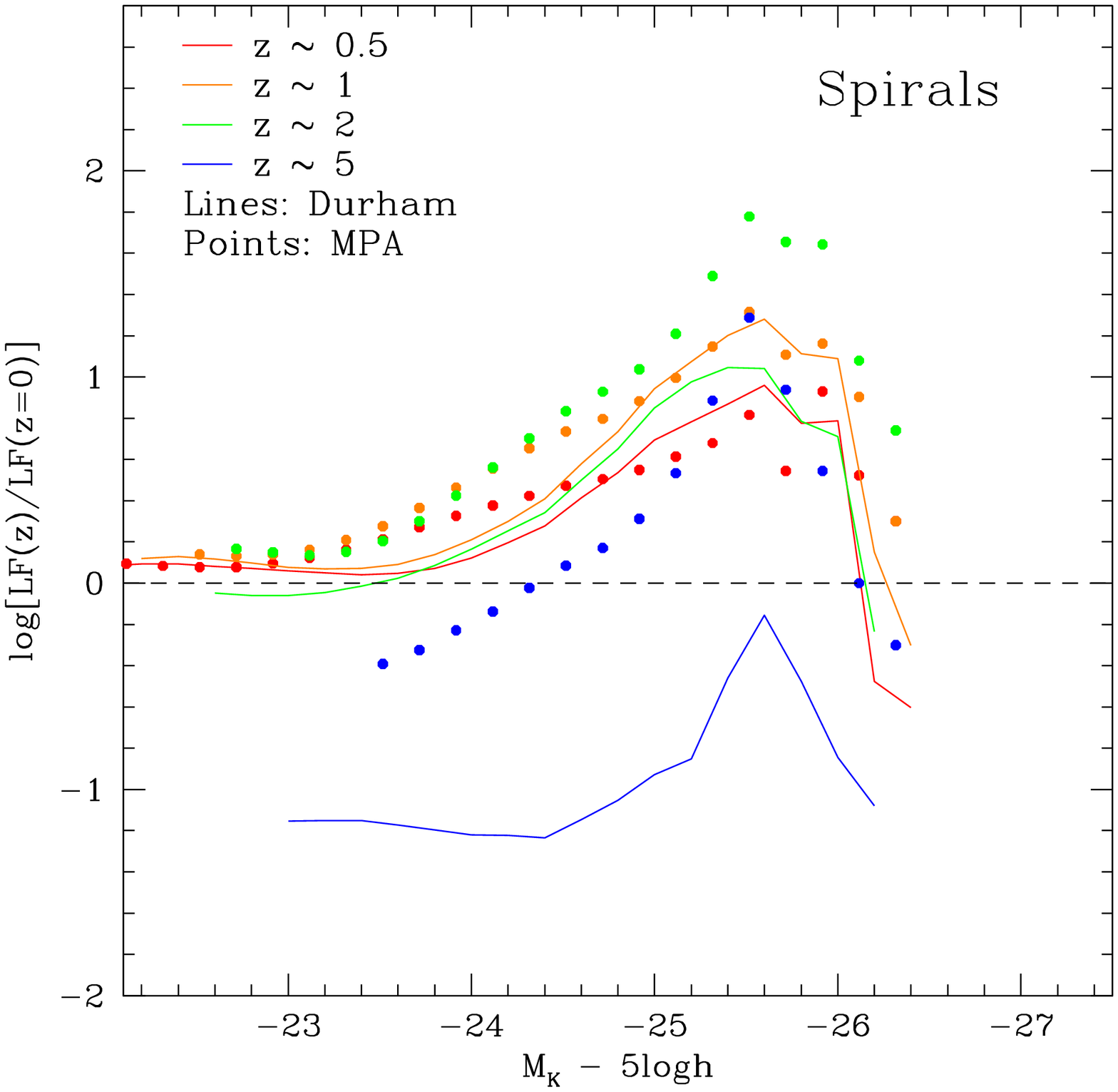} &
	\includegraphics[width=0.38\textwidth]{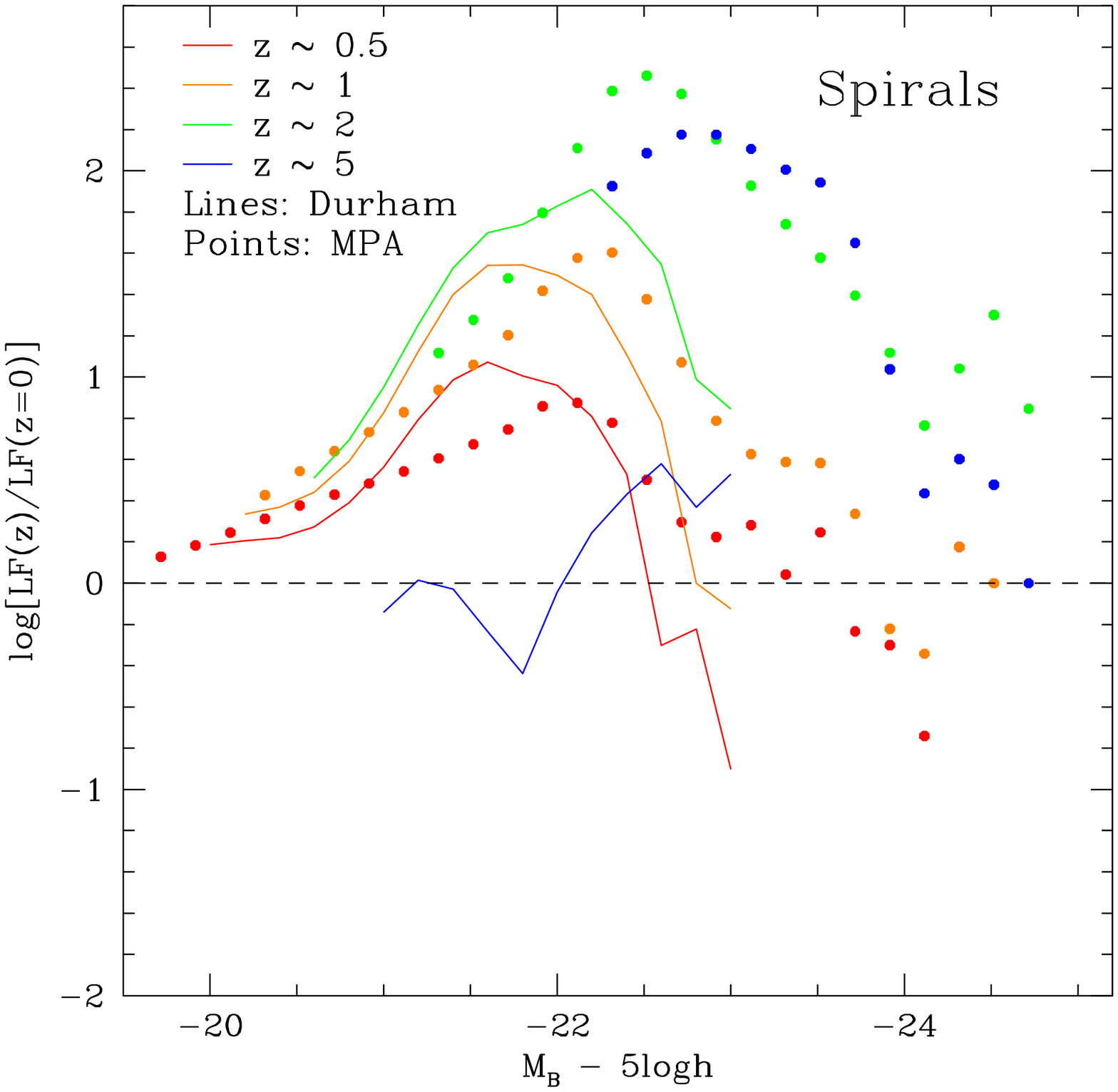}\\
	\mbox{\bf (a)} & \mbox{\bf (d)}\\
	\includegraphics[width=0.38\textwidth]{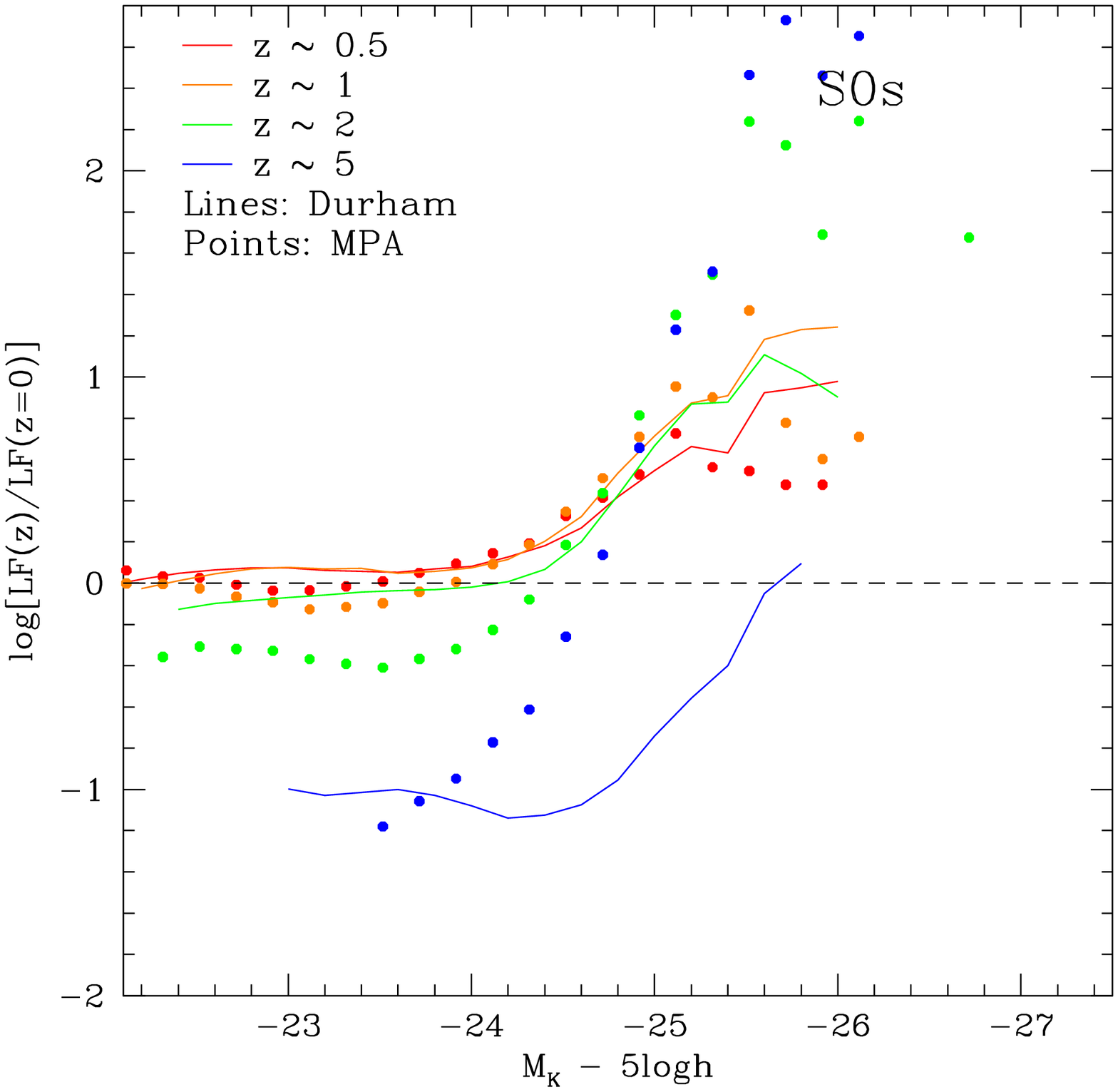} &
	\includegraphics[width=0.38\textwidth]{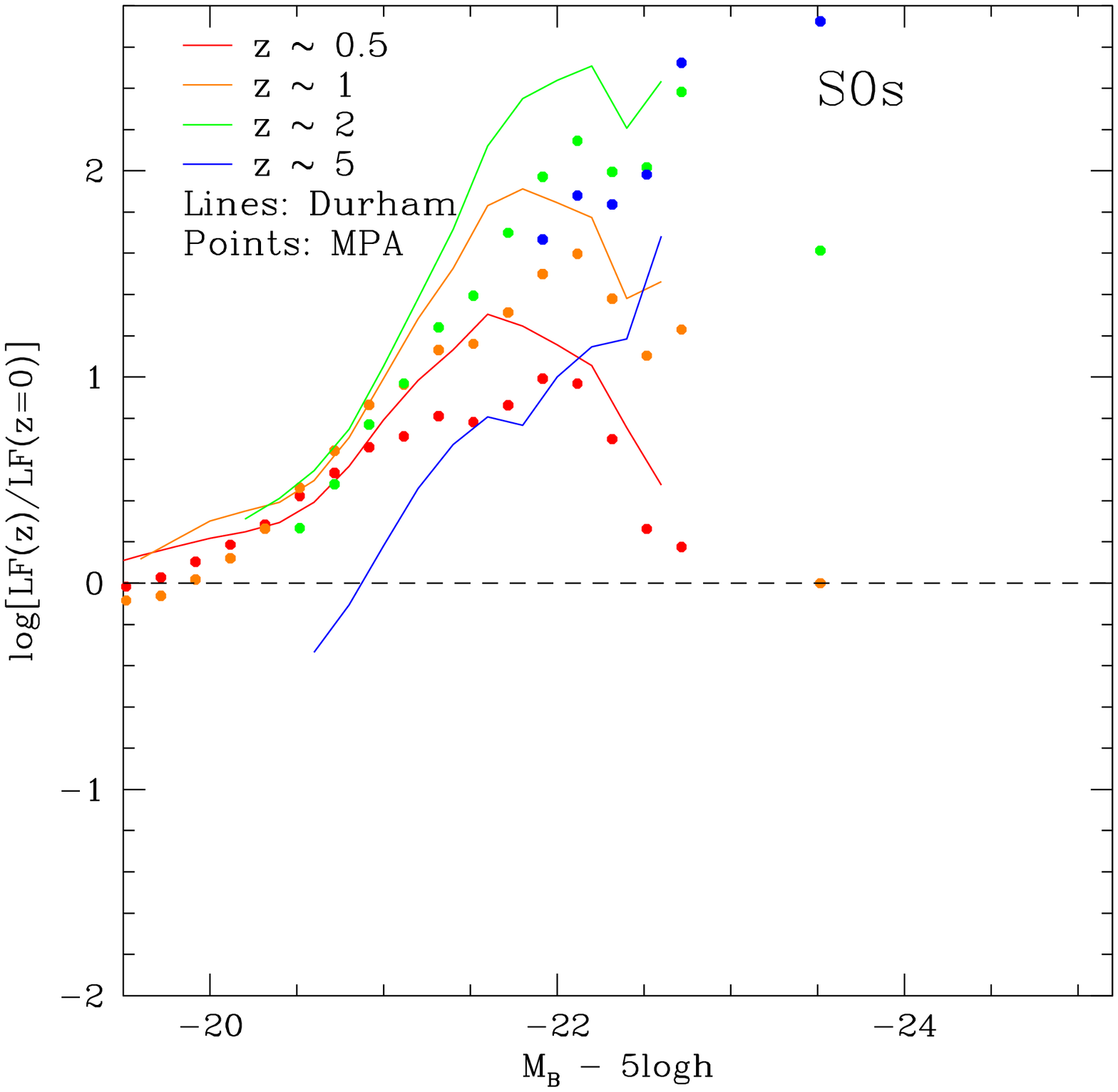}\\
\mbox{\bf (b)} & \mbox{\bf (e)}\\
	\includegraphics[width=0.38\textwidth]{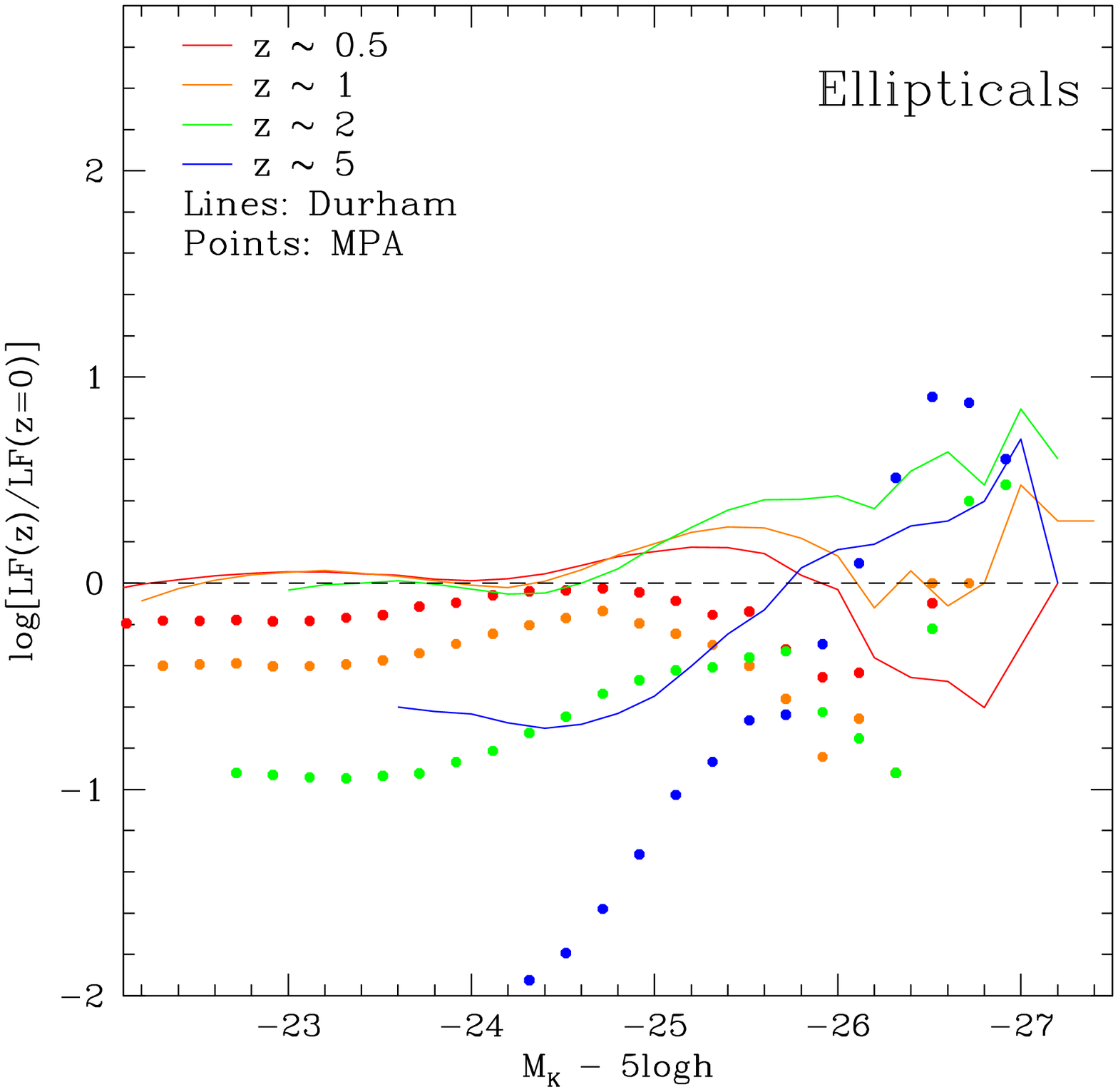} &
	\includegraphics[width=0.38\textwidth]{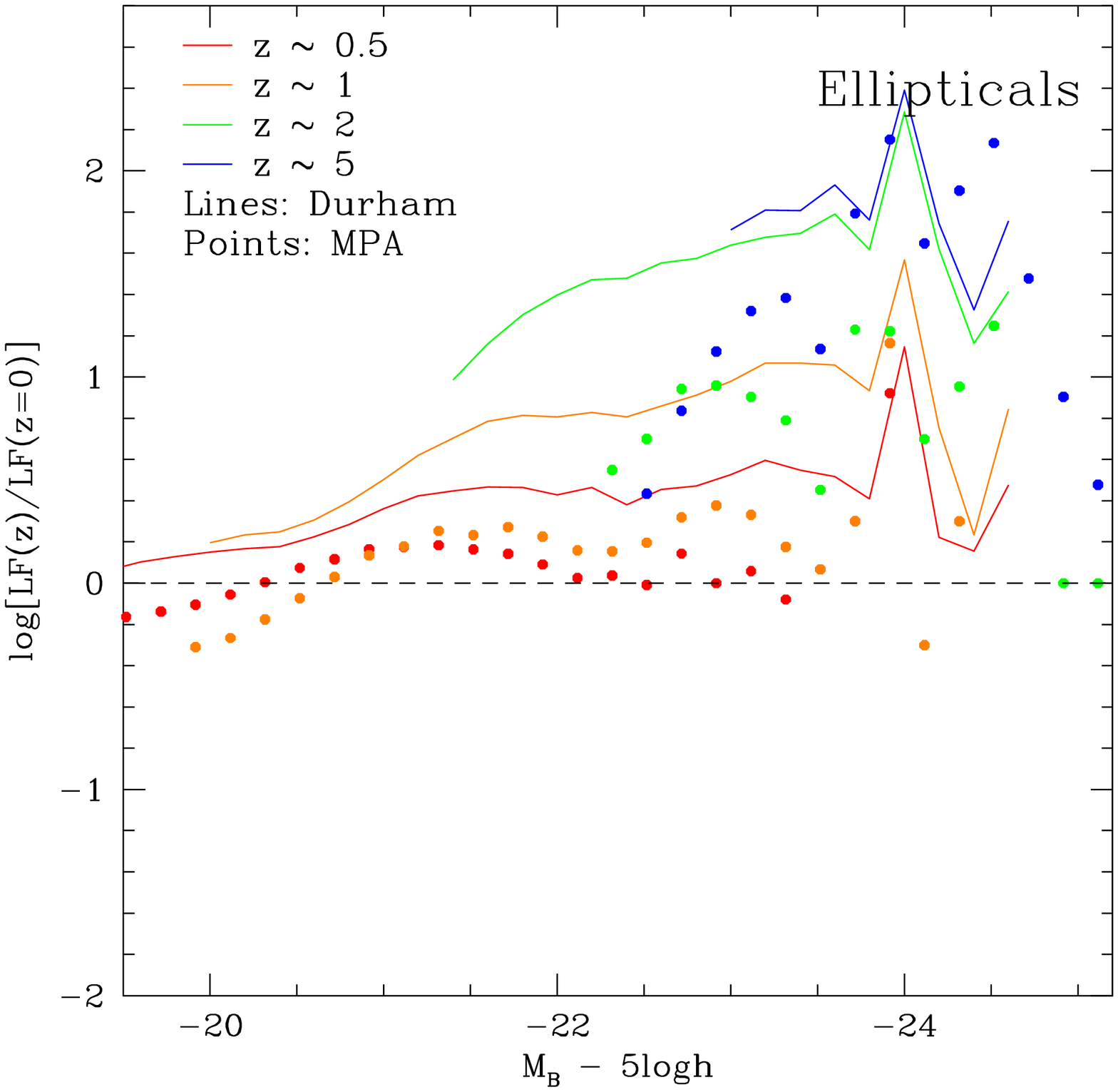}\\
\mbox{\bf (c)} & \mbox{\bf (f)}
\end{array}$
\caption{The evolution of the K-band (left column) and B-band (right column) rest-frame luminosity functions for spirals (top row), S0s (middle row) and ellipticals (bottom row), plotted as ratios to the $z=0$ luminosity functions in Fig.~\ref{fig:morphlf0}.  The curves are cut off at a magnitude faintward of which the mass cut at $4\times10^{9}h^{-1}M_{\odot}$ reduces number densities by $>5\%$.  At the bright end, they are cut off at the limit of the equivalent $z=0$ function.} 
\label{fig:morphlf}
\end{figure*}


\section{Results and discussion} \label{sec:RESULTS}

In the previous section, we saw that there is some disagreement between the Durham and MPA galaxy formation models in their predictions for the evolution of the relative abundances of galaxies of different morphological type.  In this section, we explore the reasons for these differences by investigating the formation paths of galaxies in the two models.  Firstly, we track the build up of a galaxy's dark halo and its stars with the goal of identifying how its final morphology is determined.  We then examine in more detail how spheroids form in each model and the roles played by mergers and disk instabilities.

\subsection{Halo formation}

As described in $\S$~\ref{sec:MERGE}, galaxy mergers follow the mergers of their host haloes after a time delay due to the gradual decay of the satellite's orbit through dynamical friction.  One might expect, then, that the manner of a halo's formation might play a significant role in determing the merger history of its galaxy and hence its final morphology.

We define the formation redshift of a halo, $z_{hform}$, as that at which half of its final mass was contained in a single progenitor.  Haloes in the database are linked, through their unique IDs, to the galaxies that they host, making it straightforward to relate the time of formation of a halo to the morphology of its galaxies.  We consider only `FOF haloes' since they can be associated with galaxies in both models, as opposed to the modified Durham haloes which cannot always be linked to MPA galaxies.  This allows us to compare the morphologies of galaxies in the two models for a common halo population.  We consider only central galaxies and limit our sample to haloes in the mass range $0.5-2\times10^{12}h^{-1}M_{\odot}$, a factor of two either side of the mass of a `Milky Way type' halo.

Fig.~\ref{fig:morphhaloform} shows the distribution of $z_{hform}$ for bright spirals, S0s and ellipticals in the Durham (solid line) and MPA (dashed line) models.  Relatively few galaxies in the faint sample exist in haloes within this mass range, making the equivalent plot extremely noisy.

Given that major galaxy mergers disrupt disk structures, one might na\"ively have expected later forming haloes, which merged more recently, preferentially to host ellipticals.  In the Durham model, this effect is marginal, but can be seen in the median formation redshifts for haloes hosting each type.  Halos with elliptical central galaxies have a median formation redshift of 1.17, compared to 1.28 for those with spiral or s0 centrals.  In the MPA model, haloes hosting ellipticals have a median formation redshift of 1.28, in comparison to 1.38 for those hosting S0s and 1.17 for those hosting spirals. 

The spike in haloes forming near $z=0$, seen in both models, appears to arise from the difficulty in tracking objects between snapshots.  When a halo passes through a more massive counterpart, it may appear to have merged with it, but, if it emerges on the other side, it is labelled as a newly formed object.  In this way, some merger tree branches are broken.  If the break occurs after the real formation time and on the main branch, the final time halo will appear to have formed at that epoch.  In most cases, the less massive halo falls back into its counterpart and merges before the last snapshot; thus it is no longer considered a main branch progenitor.  Hence, the spike at $z=0$ is a boundary effect - it identifies those cases where insufficient time has passed for a subsequent merger to erase evidence of a fly-through.

The above explanation is supported by the fact that the late-forming haloes are invariably found close to the virial radii of larger haloes.  Furthermore, we find that the low redshift spike disappears if we consider instead the formation of the Durham haloes, for which considerable effort has been made to repair broken merger trees and eliminate such `false mergers'~\citep{Helly2003}.

\subsection{Galaxy assembly and formation} \label{sec:a_f}

Having established the extent to which galaxy morphology depends on the epoch of halo formation, we now consider the formation of galaxies directly.  We define two different `formation epochs': i) the stellar assembly redshift, $z_{a}$, when half of the stars in a present-day galaxy were first assembled in a single progenitor, and ii) the stellar formation redshift, $z_{f}$, when half of the stars in a present-day galaxy had formed in any of its progenitors.  The three panels in Fig.~\ref{fig:a_vs_f} compare the distributions of these two redshifts for spirals, S0s and ellipticals.

\begin{figure}
\begin{center}
\leavevmode
\includegraphics[width=0.4\textwidth]{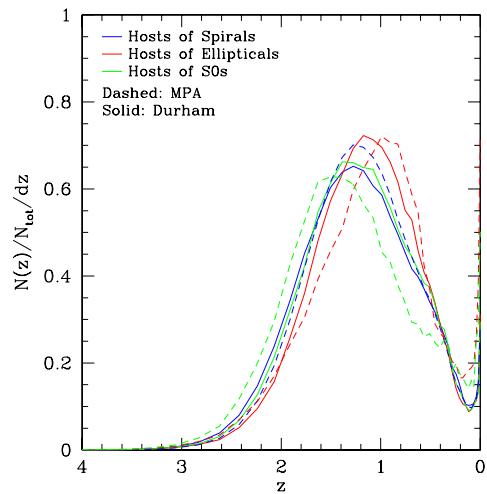}
\end{center}
\caption{The distribution of formation redshifts for haloes hosting galaxies of each morphological type in the Durham (solid) and MPA (dashed) models.  Only haloes in a `Milky Way type' mass range  ($0.5-2\times10^{12}h^{-1}M_{\odot}$) are considered here for the reasons outlined in the text.  The median formation redshifts for spiral, S0 and elliptical hosts are 1.17, 1.38 and 1.28 respectively in the MPA model, and 1.28, 1.28 and 1.17  in the Durham Model.  All curves are normalised such that the area under them is unity.}
\label{fig:morphhaloform}
\end{figure}

The two models show the same general trend for both formation and assembly to occur earliest in elliptical galaxies and latest in spirals, with S0s intermediate between the two.  Given that early-type galaxies are typically observed to have older stellar populations, it is reassuring that both models predict that they form most of their stars first.  Less intuitive, perhaps, is the fact that ellipticals also assemble their stellar mass before either S0s or spirals.  Late assembly, however, does not imply late merging.  Steady, in-place star formation in a disk naturally results in a relatively narrow distribution of assembly times, which are typically later than for elliptical and S0 galaxies where mergers play a more important role. The fact that the formation and assembly epochs of spirals (and also S0s in the Durham model) are generally coincident suggests that this is indeed the case: when the formation threshold is reached, the stars are already in one object.  In the Durham model, disks typically form at $z\sim0.55$, although a significant fraction form at $z>1$.  In the MPA model the typical formation redshift of disks is $z\sim0.9$.

\begin{figure*}
$\begin{array}{c@{\hspace{-0cm}}c@{\hspace{-0cm}}c@{\hspace{-0cm}}}
\multicolumn{1}{l}{\mbox{\bf}} &
	\multicolumn{1}{l}{\mbox{\bf}} \\
	\includegraphics[width=0.32\textwidth]{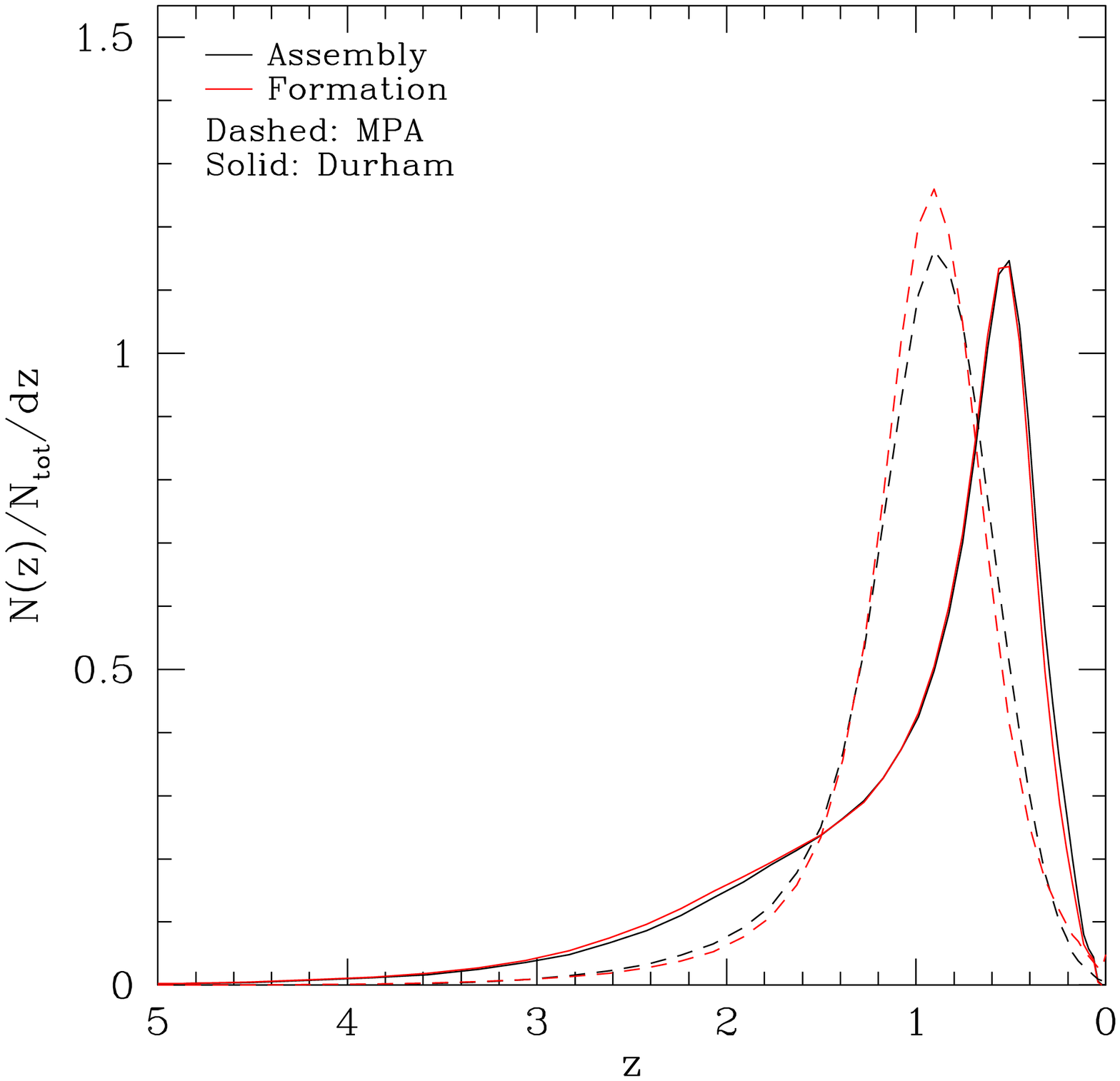} &
	\includegraphics[width=0.32\textwidth]{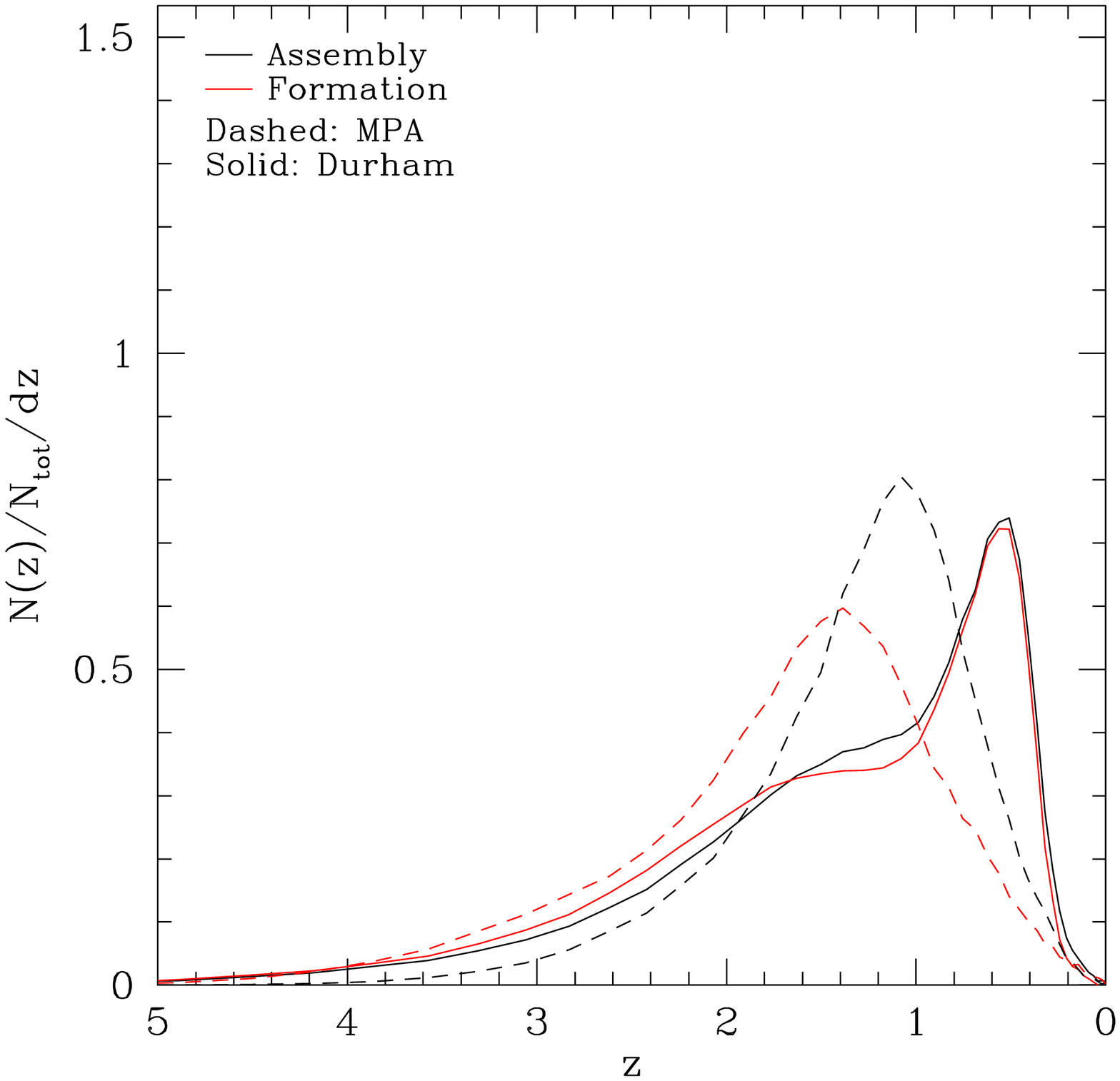} &
	\includegraphics[width=0.32\textwidth]{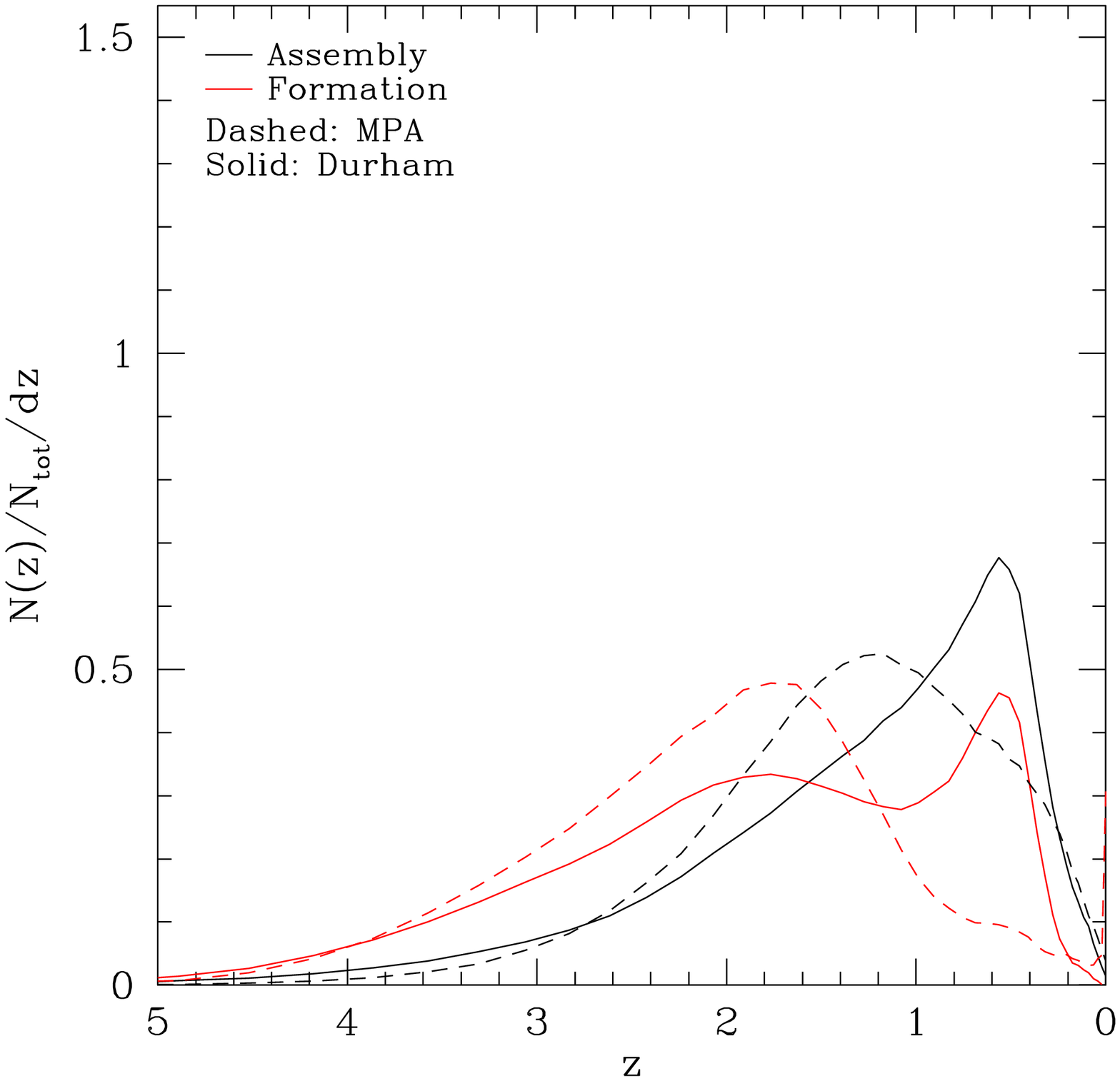} \\
	\mbox{\bf (a)} & \mbox{\bf (b)} & \mbox{\bf (c)}\\
\end{array}$
\caption{The distributions of the stellar assembly and stellar formation redshifts for bright ($M_{K}-\rm{5logh}<-22.17$) spirals \textbf{(a)}, S0s \textbf{(b)} and ellipticals \textbf{(c)} in the MPA (dashed line) and Durham (solid line) models.  All curves are normalised such that the area under them is unity.}
\label{fig:a_vs_f}
\end{figure*}

In both models, we effectively see two characteristic formation times for ellipticals and S0s, a primary peak at $z\sim1.5-2$ and a secondary peak at $z\sim0.55$.  The latter coincides with the typical epoch of spiral galaxy formation and is more pronouced in the Durham model.  This secondary peak is the result of disk instability: these ellipticals and S0s are, in effect, failed spirals, having had very similar star formation histories up to $z\sim0.55$, before becoming unstable and collapsing into a spheroid with the attendant conversion of residual gas into stars.  Some of these spheroids may partially regrow a stellar disk and end up as S0s or perhaps even as spirals, but others remain as ellipticals up to the present day, either because the instability occurred very late on, or because their gas cooling rate was restricted by feedback or lack of gas.  Since disk instabilities have a much gentler effect in the MPA model, their influence on the distribution of stellar formation times of ellipticals and S0s is minor.

Minor and major mergers play an important role in building up most of the ellipticals in the MPA model and those whose stars form in the higher redshift peak in the Durham model, as evidenced by the fact the stellar formation times are earlier than the stellar assembly times.  This asynchronism between stellar formation and assembly in CDM-based models of elliptical galaxies was first highlighted by \citet{Kauffmann1996} and has recently been investigated in the MPA model by \citet{DeLucia2006a}.  Our results for MPA ellipticals are in agreement with theirs.  The distributions of $z_{a}$ and $z_{f}$ for faint galaxies ($M_{K}-\rm{5logh}<-22.17$) in both models are similar to those for bright galaxies but shifted slightly to higher redshift and with a more significant high redshift tail.  Also, the lower redshift peak in the Durham model, corresponding to galaxies forming through disk instabilities, is much less pronounced for faint objects.

\subsection{Major mergers, minor mergers and disk instabilities}

The differences between the Durham and MPA models uncovered so far appear to stem from the roles played by the main spheroid-forming mechanisms.  In this section, we compare the relative contributions of three processes that contribute to the build-up of spheroids: major mergers, minor mergers and disk instabilities.  We remind the reader that in both models the distinction between major and minor mergers is made on the basis of the relative baryonic masses of the satellite and primary, with the boundary corresponding to $M_{sat}/M_{pri}=0.3$.  In both models, major mergers are assumed to disrupt completely any pre-existing galactic disks.

The incidence of major mergers and the redshift when the main branch progenitor of a present-day galaxy was last involved in a major merger, $z_{mm}$, can be readily obtained by tracing back the galaxy's merger tree.  The analysis below includes galaxies above our resolution limits of $1.7 \times 10^{10} h^{-1}M_{\odot}$ for the dark matter mass of the halo and $4\times10^{9}h^{-1}M_{\odot}$ for the stellar mass of the galaxy.  The statistics of major mergers are surprising and revealing.  Of the total bright $M_{K}-\rm{5logh}<-22.17$) population, only $49\%$ of ellipticals, $2\%$ of S0s and $2\%$ of spirals in the MPA model, or $41\%$, $2.5\%$ and $1.5\%$ respectively in the Durham model, undergo a main branch major merger in their entire formation history.  In both models, this fraction is largely independent of total stellar mass for spirals and S0s, but not so for ellipticals.  Combining the bright and faint populations together, in the Durham model, $100\%$ of the highest mass ($1 - 4\times10^{12}h^{-1}M_{\odot}$) ellipticals have major mergers, but this fraction falls to $<3\%$ around the lower mass limit of $4\times10^{9}h^{-1}M_{\odot}$.  In the MPA model, only $\sim15$\% of ellipticals with mass $1.6\times10^{10}h^{-1}M_{\odot}$ have a major merger, but the proportion rises sharply either side of this mass, to around $\sim80$\% at the low mass end, and close to $100\%$ for masses $>2.5\times10^{11}h^{-1}M_{\odot}$.  In both models, more than half of the total elliptical population do not undergo a main branch major merger at any point.  These results are consistent with the conclusions reached by \citet{DeLucia2006a} for MPA ellipticals.

The distribution of $z_{mm}$ for each morphological type in the bright population is displayed in Fig.~\ref{fig:mm}.  There is reasonable agreement between the two models for spirals and ellipticals, but not for the S0s, which appear to trace the elliptical distribution in the MPA case, but the spiral distribution in the Durham case.  As one might expect, ellipticals typically have had the most recent major mergers, but nonetheless, of the small fraction of spirals and S0s that have undergone major mergers at all, some experience them at very late times ($z<0.1$) and yet still manage to recover a stellar disk. The faint population shows similar behaviour to that illustrated in Fig.~\ref{fig:mm}.
Given that major mergers are this infrequent, the alternative mechanisms of minor mergers and disk instabilities must be important in spheroid formation.  Applying a similar analysis to instabilities reveals that virtually all ellipticals have experienced such an event at some point in their formation history.  However, the fraction of the total galaxy population that is unstable at a particular redshift differs substantially between the two models.  In any one snapshot, approximately $1\%$ of MPA galaxies are unstable but, in the Durham model, the figure is much higher: $\sim16\%$ at $z=6$, falling to $\sim2\%$ at $z=2$.  This difference accounts for the higher proportion of ellipticals seen at early times in the Durham model, as illustrated in Fig.~\ref{fig:morph_frac}.  Early forming disks in the Durham model are inherently less stable than those in the MPA model, largely as a consequence of their higher cold gas content.

\begin{figure}
\begin{center}
\includegraphics[width=0.4\textwidth]{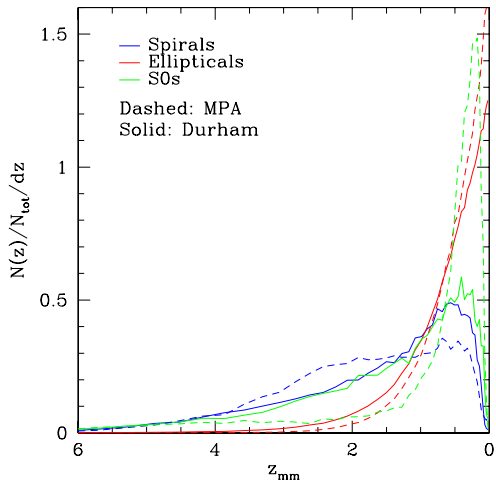}
\end{center}
\caption{The distribution of the last (main branch) major merger experienced by spirals (blue), S0s (green) and ellipticals (red) in the Durham (solid line) and MPA (dashed line) bright ($M_{K}-\rm{5logh}<-22.17$) galaxy populations.  Each curve is normalised such that the area under it is unity.  The median values of $z_{mm}$ are 2.62, 0.83, and 0.41 for spirals, S0s and ellipticals respectively in the MPA model, and 2.07, 1.63 and 0.51 in the Durham model.}
\label{fig:mm}
\end{figure}

To quantify the relative impact of instabilities, major mergers and minor mergers on galaxy morphology, we determine the fraction of stellar mass in present-day spheroids that was incorporated into the spheroid by each process.  The complexity of the SQL query needed for this calculation made it necessary to use the `milli-millennium' datasets, which were generated using the same two semi-analytic models on halo merger trees constructed from a smaller version of the Millennium simulation (1/512 of the volume).  Even though the milli-millennium simulation has somewhat less structure on the largest scales, its cosmological parameters and other attributes are identical, so our statistical results will be similar in both simulations.

\begin{figure}
\begin{center}
\includegraphics[width=0.4\textwidth]{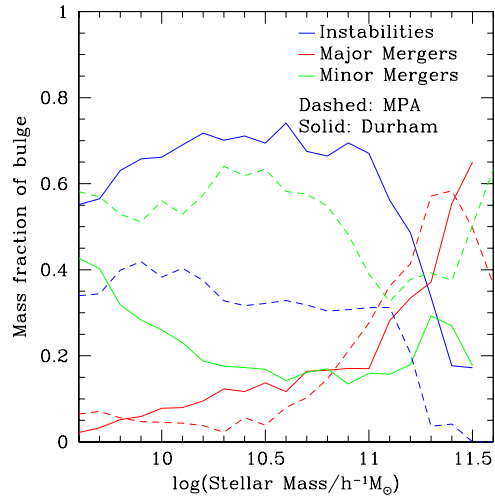}
\end{center}
\caption{The fractional contributions of instabilities (blue), minor mergers (green) and major mergers (red) to the stellar content of spheroids, as a function of total galaxy mass, in the Durham (solid lines) and MPA (dashed lines) models.}
\label{fig:all_frac}
\end{figure}

Fig.~\ref{fig:all_frac} shows the average fraction of stellar mass in the bulge component of a galaxy that is present due to each of the three spheroid-forming mechanisms, as a function of the total stellar mass of the galaxy.  These fractions include stars that may have formed in the bulge as a result of a starburst induced by any of the three processes.  As expected from our previous results, instabilities contribute considerably more to bulges in the Durham model, dominating in all but the most massive galaxies, where most stars are brought in by, or formed in, major mergers.  In the MPA model, it is minor mergers that are responsible for the bulk of stellar mass in spheroids, but, again, major mergers are dominant at the highest masses. 

Other galaxy formation models have used slightly different values for and definitions of the mass ratio that identifies a major merger.  For example \citet{Somerville2008} use a value of 0.25 and include all dark matter within a radius of twice the NFW scale-length of the halo. We have tested the sensitivity of the above result to the baryonic mass ratio used to define a major merger in the Durham model.  Values between 0.25 and 0.35 modify the fractions in Fig.~\ref{fig:all_frac} by only a few percent and a value of $<0.01$ is required in order to make major mergers the dominant channel for bulge growth above a stellar mass of $\sim5\times10^{10}h^{-1}M_\odot$.  Considering instead a versions of the Durham model which employs a more realistic prescription for gas stripping in satellites \citep{Font2008}, we find that major mergers occur more frequently.  This appears to be mainly the result of a higher satellite mass (in cold baryons) at the point of merger, which reduces the mass ratio.   Also, since the total baryonic satellite mass is larger, the merger timescale is shorter, which in turn means that the central galaxy has less time to increase the mass ratio by cooling more gas before the merger occurs.  Nonetheless, in this model, major mergers still account for $<20\%$ of bulge stars in galaxies less massive than $\sim5\times10^{11}h^{-1}M_\odot$ and at higher masses they contribute almost identical fractions to those seen in the Bower \rm{et al.} model.

Several authors have considered the role of mergers in spheroid formation, although none to our knowledge has included disk instabilities alongside mergers as a channel for bulge growth. \citet{Kang2007} used a semi-analytic model similar in structure to the Durham and MPA models and were able to match the observed fraction of boxy early-type galaxies as a function of luminosity and halo mass by assuming that gas poor major mergers result in boxy remnants.  \citet{Khochfar2006} used another semi-analytic model, though without AGN feedback or N-body dark matter merger trees, to examine the contribution of `quiescent'and `merger' processes - stars formed in progenitor disks and in merger-induced starbursts respectively.  They found that bulges and elliptical galaxies are dominated by quiescently formed stars.  The aspects of spheroid formation investigated by \citet{Khochfar2006} and \citet{Kang2007} and their methods of analysis differ considerably from those presented here, making it difficult to determine whether their results are consistent or not with spheroid formation in the Durham and MPA models.  \citet{Hopkins2008} used a `semi-empirical' model, based on an observationally constrained halo occupation distribution, combined with theoretical estimates for merger timescales to track galaxy mergers.  They found that gas rich major mergers can account for the observed mass-density of all red, early types at $z=0$ and the dichotomy in the kinematic properties of ellipticals.   These results appear, qualitatively at least, to be inconsistent with the Durham and MPA models, suggesting that major mergers alone are sufficient to explain the formation of spheroids.  The frequency of major mergers implied by their results certainly seems to exceed that in either the Durham or MPA semi-analytic prescriptions, though we note that they do not attempt to follow the processes of cooling, star formation and feedback that influence the major merger rates in these two models.

As the above results have demonstrated, not only are major mergers a relatively infrequent occurrence in both the Durham and MPA models, but their overall fractional contribution to building spheroids is typically small compared to the other two mechanisms in all but the largest galaxies.  The importance of disk instabilities, particularly in the Durham model, suggests a possible observational test of this, perhaps unexpected, conclusion, because the rotation properties of spheroids should differ depending on whether they formed predominantly through mergers or instabilities.

The relative importance of rotational and random motions in spheroidal systems is measured by the ratio of rotational speed to velocity dispersion, $V/\sigma$, which is often plotted against the galaxy's ellipticity to determine whether its shape is supported by rotation or by velocity anisotropy \citep{Illingworth1977,Binney1978,Kormendy1993,Kormendy2004,Kormendy2008}.  During major mergers, tidal forces transfer angular momentum from the baryonic component to the dark halo~\citep[e.g.,][]{Frenk1985,Zavala2008} so the resulting merger remnants tend to have low $V/\sigma$ and a shape that is supported by velocity anisotropy.  Spheroids forming through disk instabilities on the other hand are expected to retain a significant proportion of the angular momentum of the progenitor disk, ending up close to the `isotropic rotator' line in the $V/\sigma$ {\em vs} flattening diagram.  Our findings are qualitatively consistent with observations: low luminosity ellipticals \citep{Davies1983} and most  `classical' spiral bulges \citep{Illingworth1982,Kormendy1982} are found to  be nearly isotropic, oblate rotators, whereas most giant ellipticals have  insignificant rotation \citep{Bertola1975,Illingworth1977}.  A detailed  analysis of fast and slow rotators using integral field spectroscopy  \citep{Emsellem2007} leads to similar conclusions, with massive ellipticals  exhibiting negligible rotation and the fastest rotators tending to be of lower luminosity.

An important caveat to the test described above is that major mergers can also produce spheroids that are disky and rapidly rotating.  Several simulations of mergers between equal mass, gas rich disks have shown that, despite a large burst taking place, enough gas may survive to form a new, rotationally supported stellar disk within the remnant \citep{Barnes1996,Bournaud2005,Springel2005b,Naab2006,Hopkins2008}.  Merger geometry is also likely to influence remnant structure \citep{Hernquist1995,Bournaud2005}.  These considerations may make our proposed test of the models less clear cut. 

The differences in the fractions of spheroid stars added by instabilities in the two models are unsurprising given the contrasting prescriptions for instability already discussed, but it is not simply the dominance of this mechanism that decreases the fractional contribution from minor mergers in the Durham case.  The average mass added by minor mergers is typically somewhat larger in the MPA model for any given galaxy.  This difference appears to stem from two aspects of the models.  The Durham model has a longer characteristic star formation timescale and also stronger supernovae feedback, which inhibits star formation in lower mass galaxies and efficiently depletes the cold gas available to form stars in satellites.  Hence, there is less mass on average in each Durham satellite.  At the high mass end, however, supernova feedback is less effective, and so the mass brought in by major mergers is more similar in the two models. As Fig.~\ref{fig:all_frac} shows, the two models, in fact, agree rather well on the fractional contribution of major mergers to spheroid formation.

\section{Conclusions} \label{sec:CONCLUSIONS}

We have investigated the origin and evolution of galaxy morphology in the $\Lambda$CDM cosmology using two different and publicly available semi-analytic galaxy formation models, the Durham and MPA models \citep{Bower2006,DeLuciaBlaizot2007}, both based on the Millennium dark matter simulation.  Morphology was defined for each galaxy by means of the bulge-to-total ratio in K-band luminosity; we distinguish between the behaviour of bright ($M_{K}-\rm{5logh}<-22.17$) and faint ($M_{K}-\rm{5logh}>-22.17$) galaxies.

There are many similarities, but also fundamental disagreements in the predictions of the two models.  One of the largest differences involves the redshift evolution of the morphological fractions: spirals are the most common systems at all times in the MPA model, while ellipticals dominate in the Durham model at redshifts $z>4$. The main cause of this and other differences highlighted below is the different treatment of disk instabilities in the two models.  In the Durham model, instabilities are not only more common at early times, but also have more catastrophic consquences: the sudden conversion of the entire disk, first into a bar and then into a spheroid.  Nevertheless, the two models predict a fairly comparable morphological mix at the present day.

In comparison with data from \citet{Benson2007}, the substantial systematic uncertainties involved in the process of fitting bulge and disk light profiles mean that only weak constraints can be placed on the fraction of each morphological type as a function of absolute r-band magnitude.  Neither model is obviously inconsistent with the data, but clearly smaller systematic uncertainties would be desirable.  The bright galaxy populations consist of $51\%$, $15\%$ and $34\%$ of spirals, S0s and ellipticals respectively in the Durham model, and $67\%$, $8\%$ and $25\%$ respectively in the MPA model.  

For typical Milky-Way type haloes, we find that the formation time of the halo has, at most, only a small effect on the morphology of the central galaxy.  In both models, haloes hosting ellipticals form slightly earlier than haloes hosting spirals or S0s, although this effect is only marginal in the Durham model.
 
Our analysis confirms previous studies \citep{Kauffmann1996,DeLucia2006a} that show that ellipticals galaxies form most of their stars before the bulk of the galaxy is assembled.  In fact, the ellipticals form and assemble their stars earlier than galaxies of other present-day morphological types, although the late `assembly' times of spirals are a consequence of in-situ star formation, rather than mergers.  In general, the MPA model predicts slightly earlier star formation for galaxies of all types.  In the Durham model, there is a significant population of S0s and ellipticals that formed at relatively low redshift from the recent collapse of unstable disks, yielding spheroids that never regrow a sizeable stellar disk.  The differences in the stellar formation histories predicted by the two models are, in principle, accessible to observational tests.

Perhaps the most surprising conclusion of this study is that, with the exception of the brightest ellipticals (stellar mass $M_*\gsim 2.5\times 10^{11}h^{-1}M_\odot$), major mergers are {\it not} the primary mechanism by which most spheroids (ellipticals and spiral bulges) assemble their mass.  Elliptical galaxies are more likely than spirals or S0s to have undergone a recent major merger, but nonetheless more than half of them do not experience one throughout their whole formation history (involving galaxies above the resolution limits of our calculations).  Major mergers are even rarer for spirals, affecting $<4\%$ of galaxies; their bulges are almost universally formed by minor mergers or disk instabilities.  These conclusions would appear to be quite robust for galaxy formation in a $\Lambda$CDM universe since they are common to both the MPA and Durham models, in spite of the substantial differences in the detailed modelling of key physical processes.

An important difference between the two models concerns the channel through which less massive ellipticals and bulges acquire their mass.  In the Durham model, disk instabilities are responsible for the bulk of the stellar mass in spheroids at $z=0$, whereas in the MPA model, minor mergers contribute the most.  In the Durham model at least, these results appear to be relatively insensitive to the mass ratio at which a merger is defined as `major'.  Varying the value between 0.25 and 0.35 typically alters the fractions in fig. \ref{fig:all_frac} by less than a few percent for any given stellar mass; in order for major mergers to become the dominant channel for bulge growth in galaxies larger than $\sim5\times10^{10}h^{-1}M_\odot$, the ratio would need to be lowered to $<0.01$.  Adding a more realistic prescription for gas stripping of satellites \citep{Font2008} increases the importance of major mergers, but they still contribute $<20\%$ of spheroid mass in galaxies smaller than $\sim5\times10^{11}h^{-1}M_\odot$.  Mergers as a whole contribute less mass to spheroids in the Durham model because strong supernova feedback suppresses star formation especially strongly in low mass satellite galaxies.  

The dominant role of disk instabilities in the production of spheroids, particularly in the Durham model, offers a possible observational test, since spheroids formed this way will tend to rotate faster than spheroids formed through mergers.  However, this test would be complicated by the fact that gas rich major mergers can also produce high $V/\sigma$ remnants.  What is clear is that a more precise comparison between theory and observations will require more detailed modelling of disk instabilities than we have attempted here.

More generally, semi-analytic techniques offer, at present, only a somewhat crude way to model the processes that establish galaxy morphology.  It is therefore remarkable that such a simplified approach can lead to some important conclusions that appear to be robust to the modelling details.  This approach also serves to identify specific processes that deserve further study, for example, through direct numerical simulation.


\section*{Acknowledgments}

We are grateful to John Helly and Gerard Lemson for their technical support with the Millennium Run data catalogues in Durham and Garching respectively and to Richard Bower and Simon White for constructive discussions and advice.  We would also like to thank Gabriella De Lucia for assistance with our analysis of the MPA model and Andrew Benson for providing the B-T data plotted in Fig. \ref{fig:BT}.  CSF acknowldeges the Aspen Center for Physics for their unique hospitality while parts of this paper were being written.  Finally, we thank the anonymous referee for many helpful suggestions which have substantially improved the manuscript.  OHP acknowledges an STFC PhD studentship, VRE a Royal Society University Research Fellowship and CSF a Royal Society Wolfson Research Merit award.  This work was supported also by an STFC rolling grant to the ICC. The Millennium simulation was carried out by the Virgo consortium and the database used in this paper, together with the web application that provides online access to it, were constructed as part of the activities of the German Astrophysical Virtual Observatory.

\begin{thebibliography}{}

\bibitem[\protect\citeauthoryear{{Abadi}, {Moore} \& {Bower}}{{Abadi}
  et~al.}{1999}]{Abadi1999}
{Abadi} M.~G.,  {Moore} B.,    {Bower} R.~G.,  1999, \mnras, 308, 947

\bibitem[\protect\citeauthoryear{{Barnes}}{{Barnes}}{1992}]{Barnes1992}
{Barnes} J.~E.,  1992, \apj, 393, 484

\bibitem[\protect\citeauthoryear{{Barnes} \& {Hernquist}}{{Barnes} \&
  {Hernquist}}{1996}]{Barnes1996}
{Barnes} J.~E.,  {Hernquist} L.,  1996, \apj, 471, 115

\bibitem[\protect\citeauthoryear{{Bekki} \& {Shioya}}{{Bekki} \&
  {Shioya}}{1997}]{Bekki1997}
{Bekki} K.,  {Shioya} Y.,  1997, \apj, 478, L17

\bibitem[\protect\citeauthoryear{{Bell}}{{Bell}}{2008}]{Bell2008}
{Bell} E.~F.,  2008, ArXiv e-prints, astro-ph/0801.0599

\bibitem[\protect\citeauthoryear{{Bell} \& {de Jong}}{{Bell} \& {de
  Jong}}{2001}]{Bell2001}
{Bell} E.~F.,  {de Jong} R.~S.,  2001, \apj, 550, 212

\bibitem[\protect\citeauthoryear{{Benson}}{{Benson}}{2005}]{Benson2005}
{Benson} A.~J.,  2005, \mnras, 358, 551

\bibitem[\protect\citeauthoryear{{Benson}, {Dzanovic}, {Frenk} \&
  {Sharples}}{{Benson} et~al.}{2007}]{Benson2007}
{Benson} A.~J.,  {Dzanovic} D.,  {Frenk} C.~S.,    {Sharples} R.,  2007,
  \mnras, 379, 841

\bibitem[\protect\citeauthoryear{{Bertola} \& {Capaccioli}}{{Bertola} \&
  {Capaccioli}}{1975}]{Bertola1975}
{Bertola} F.,  {Capaccioli} M.,  1975, \apj, 200, 439

\bibitem[\protect\citeauthoryear{{Binney}}{{Binney}}{1978}]{Binney1978}
{Binney} J.,  1978, \mnras, 183, 501

\bibitem[\protect\citeauthoryear{{Blanton}, {Dalcanton}, {Eisenstein},
  {Loveday}, {Strauss}, {SubbaRao} \& {Weinberg}}{{Blanton}
  et~al.}{2001}]{Blanton2001}
{Blanton} M.~R.,  {Dalcanton} J.,  {Eisenstein} D.,  {Loveday} J.,  {Strauss}
  M.~A.,  {SubbaRao} M.,    {Weinberg} D.~H.,  2001, \aj, 121, 2358

\bibitem[\protect\citeauthoryear{{Bournaud}, {Jog} \& {Combes}}{{Bournaud}
  et~al.}{2005}]{Bournaud2005}
{Bournaud} F.,  {Jog} C.~J.,    {Combes} F.,  2005, \aap, 437, 69

\bibitem[\protect\citeauthoryear{{Bower}, {Benson}, {Malbon}, {Helly}, {Frenk},
  {Baugh}, {Cole} \& {Lacey}}{{Bower} et~al.}{2006}]{Bower2006}
{Bower} R.~G. {\etal},  2006, \mnras, 370, 645

\bibitem[\protect\citeauthoryear{{Brinchmann} \& {Ellis}}{{Brinchmann} \&
  {Ellis}}{2000}]{Brinchmann2000}
{Brinchmann} J.,  {Ellis} R.~S.,  2000, \apj, 536, L77

\bibitem[\protect\citeauthoryear{{Bruzual} \& {Charlot}}{{Bruzual} \&
  {Charlot}}{2003}]{Bruzual2003}
{Bruzual} G.,  {Charlot} S.,  2003, \mnras, 344, 1000

\bibitem[\protect\citeauthoryear{{Chabrier}}{{Chabrier}}{2003}]{Chabrier2003}
{Chabrier} G.,  2003, \pasp, 115, 763

\bibitem[\protect\citeauthoryear{{Cole}, {Lacey}, {Baugh} \& {Frenk}}{{Cole}
  et~al.}{2000}]{Cole2000}
{Cole} S.,  {Lacey} C.~G.,  {Baugh} C.~M.,    {Frenk} C.~S.,  2000, \mnras,
  319, 168

\bibitem[\protect\citeauthoryear{{Cole}, {Norberg}, {Baugh}, {Frenk},
  {Bland-Hawthorn}, {Bridges} \& {Cannon}}{{Cole} et~al.}{2001}]{Cole01}
{Cole} S. {\etal},  2001, \mnras, 326, 255

\bibitem[\protect\citeauthoryear{{Colless}, {Dalton}, {Maddox}, {Sutherland},
  {Norberg} \& {Cole}}{{Colless} et~al.}{2001}]{Colless2001}
{Colless} M. {\etal},  2001, \mnras, 328, 1039

\bibitem[\protect\citeauthoryear{{Cox}, {Dutta}, {Di Matteo}, {Hernquist},
  {Hopkins}, {Robertson} \& {Springel}}{{Cox} et~al.}{2006}]{Cox2006}
{Cox} T.~J.,  {Dutta} S.~N.,  {Di Matteo} T.,  {Hernquist} L.,  {Hopkins}
  P.~F.,  {Robertson} B.,    {Springel} V.,  2006, \apj, 650, 791

\bibitem[\protect\citeauthoryear{{Croft}, {Di Matteo}, {Springel} \&
  {Hernquist}}{{Croft} et~al.}{2008}]{Croft2008}
{Croft} R. A.~C.,  {Di Matteo} T.,  {Springel} V.,    {Hernquist} L.,  2008,
  ArXiv e-prints, astro-ph/0803.4003

\bibitem[\protect\citeauthoryear{{Croton}, {Springel}, {White}, {De Lucia},
  {Frenk} \& {Gao}}{{Croton} et~al.}{2006}]{Croton2006}
{Croton} D.~J. {\etal},  2006, \mnras, 365, 11

\bibitem[\protect\citeauthoryear{{Daddi}}{{Daddi}}{2004}]{Daddi2004}
{Daddi} E.,  2004, \apj, 600, L127

\bibitem[\protect\citeauthoryear{{Davies}, {Efstathiou}, {Fall}, {Illingworth}
  \& {Schechter}}{{Davies} et~al.}{1983}]{Davies1983}
{Davies} R.~L.,  {Efstathiou} G.,  {Fall} S.~M.,  {Illingworth} G.,
  {Schechter} P.~L.,  1983, \apj, 266, 41

\bibitem[\protect\citeauthoryear{{Davis}, {Efstathiou}, {Frenk} \&
  {White}}{{Davis} et~al.}{1985}]{Davis1985}
{Davis} M.,  {Efstathiou} G.,  {Frenk} C.~S.,    {White} S. D.~M.,  1985, \apj,
  292, 371

\bibitem[\protect\citeauthoryear{{De Lucia} \& {Blaizot}}{{De Lucia} \&
  {Blaizot}}{2007}]{DeLuciaBlaizot2007}
{De Lucia} G.,  {Blaizot} J.,  2007, \mnras, 375, 2

\bibitem[\protect\citeauthoryear{{De Lucia}, {Springel}, {White}, {Croton} \&
  {Kauffmann}}{{De Lucia} et~al.}{2006}]{DeLucia2006a}
{De Lucia} G.,  {Springel} V.,  {White} S. D.~M.,  {Croton} D.,    {Kauffmann}
  G.,  2006, \mnras, 366, 499

\bibitem[\protect\citeauthoryear{{de Vaucouleurs}}{{de
  Vaucouleurs}}{1959}]{DV1959}
{de Vaucouleurs} G.,  1959, Handbuch der Physik, 53, 275

\bibitem[\protect\citeauthoryear{{Dickinson}}{{Dickinson}}{2000}]{Dickinson200%
0}
{Dickinson} M.,  2000, in Astronomy, physics and chemistry of {$H^+_3$}
  Vol.~358 of {Royal Society of London Philosophical Transactions Series A},
  {The first galaxies: structure and stellar populations}.
pp 2001--+

\bibitem[\protect\citeauthoryear{{Dressler}}{{Dressler}}{1980}]{Dressler1980}
{Dressler} A.,  1980, \apj, 236, 351

\bibitem[\protect\citeauthoryear{{Dressler}, {Oemler}, {Couch}, {Smail},
  {Ellis} \& {Barger}}{{Dressler} et~al.}{1997}]{Dressler1997}
{Dressler} A. {\etal},  1997, \apj, 490, 577

\bibitem[\protect\citeauthoryear{{Driver}, {Popescu}, {Tuffs}, {Liske},
  {Graham}, {Allen} \& {de Propris}}{{Driver} et~al.}{2007}]{Driver2007}
{Driver} S.~P.,  {Popescu} C.~C.,  {Tuffs} R.~J.,  {Liske} J.,  {Graham} A.~W.,
   {Allen} P.~D.,    {de Propris} R.,  2007, \mnras, 379, 1022

\bibitem[\protect\citeauthoryear{{Efstathiou}, {Lake} \&
  {Negroponte}}{{Efstathiou} et~al.}{1982}]{Efstathiou1982}
{Efstathiou} G.,  {Lake} G.,    {Negroponte} J.,  1982, \mnras, 199, 1069

\bibitem[\protect\citeauthoryear{{Emsellem}, {Cappellari}, {Krajnovi{\'c}},
  {van de Ven}, {Bacon} \& {Bureau}}{{Emsellem} et~al.}{2007}]{Emsellem2007}
{Emsellem} E. {\etal},  2007, \mnras, 379, 401

\bibitem[\protect\citeauthoryear{{Ferrarese} \& {Merritt}}{{Ferrarese} \&
  {Merritt}}{2000}]{Ferrarese2000}
{Ferrarese} L.,  {Merritt} D.,  2000, \apj, 539, L9

\bibitem[\protect\citeauthoryear{{Font}, {Bower}, {McCarthy}, {Benson},
  {Frenk}, {Helly}, {Lacey}, {Baugh} \& {Cole}}{{Font} et~al.}{2008}]{Font2008}
{Font} A.~S. {\etal},  2008, \mnras, 389, 1619

\bibitem[\protect\citeauthoryear{{Frenk}, {White}, {Efstathiou} \&
  {Davis}}{{Frenk} et~al.}{1985}]{Frenk1985}
{Frenk} C.~S.,  {White} S. D.~M.,  {Efstathiou} G.,    {Davis} M.,  1985, \nat,
  317, 595

\bibitem[\protect\citeauthoryear{{Gadotti}}{{Gadotti}}{2008}]{Gadotti2008}
{Gadotti} D.~A.,  2008, ArXiv e-prints

\bibitem[\protect\citeauthoryear{{Giavalisco}, {Steidel} \&
  {Macchetto}}{{Giavalisco} et~al.}{1996}]{Giavalisco1996}
{Giavalisco} M.,  {Steidel} C.~C.,    {Macchetto} F.~D.,  1996, \apj, 470, 189

\bibitem[\protect\citeauthoryear{{Gunn} \& {Gott}}{{Gunn} \&
  {Gott}}{1972}]{GunnGott1972}
{Gunn} J.~E.,  {Gott} J.~R.,  1972, \apj, 176, 1

\bibitem[\protect\citeauthoryear{{H{\"a}ring-Neumayer}, {Cappellari}, {Rix},
  {Hartung}, {Prieto}, {Meisenheimer} \& {Lenzen}}{{H{\"a}ring-Neumayer}
  et~al.}{2006}]{Haring2006}
{H{\"a}ring-Neumayer} N.,  {Cappellari} M.,  {Rix} H.~W.,  {Hartung} M.,
  {Prieto} M.~A.,  {Meisenheimer} K.,    {Lenzen} R.,  2006, \apj, 643, 226

\bibitem[\protect\citeauthoryear{{Harker}, {Cole}, {Helly}, {Frenk} \&
  {Jenkins}}{{Harker} et~al.}{2006}]{Harker2006}
{Harker} G.,  {Cole} S.,  {Helly} J.,  {Frenk} C.~S.,    {Jenkins} A.,  2006,
  \mnras, 367, 1039

\bibitem[\protect\citeauthoryear{{Helly}, {Cole}, {Frenk}, {Baugh}, {Benson} \&
  {Lacey}}{{Helly} et~al.}{2003}]{Helly2003}
{Helly} J.~C.,  {Cole} S.,  {Frenk} C.~S.,  {Baugh} C.~M.,  {Benson} A.,
  {Lacey} C.,  2003, \mnras, 338, 903

\bibitem[\protect\citeauthoryear{{Hernquist} \& {Mihos}}{{Hernquist} \&
  {Mihos}}{1995}]{Hernquist1995}
{Hernquist} L.,  {Mihos} J.~C.,  1995, \apj, 448, 41

\bibitem[\protect\citeauthoryear{{Hopkins}, {Cox}, {Younger} \&
  {Hernquist}}{{Hopkins} et~al.}{2008}]{Hopkins2008}
{Hopkins} P.~F.,  {Cox} T.~J.,  {Younger} J.~D.,    {Hernquist} L.,  2008,
  ArXiv e-prints

\bibitem[\protect\citeauthoryear{{Hoyle}}{{Hoyle}}{1949}]{Hoyle1949}
{Hoyle} F.,  1949, Ohio: Central Air Documents Office, pp 195--197

\bibitem[\protect\citeauthoryear{{Hubble}}{{Hubble}}{1926}]{Hubble1926}
{Hubble} E.~P.,  1926, \apj, 64, 321

\bibitem[\protect\citeauthoryear{{Hubble}}{{Hubble}}{1936}]{Hubble1936}
{Hubble} E.~P.,  1936, Yale University Press

\bibitem[\protect\citeauthoryear{{Illingworth}}{{Illingworth}}{1977}]{Illingwo%
rth1977}
{Illingworth} G.,  1977, \apjl, 218, L43

\bibitem[\protect\citeauthoryear{{Illingworth} \& {Schechter}}{{Illingworth} \&
  {Schechter}}{1982}]{Illingworth1982}
{Illingworth} G.,  {Schechter} P.~L.,  1982, \apj, 256, 481

\bibitem[\protect\citeauthoryear{{Kang}, {van den Bosch} \& {Pasquali}}{{Kang}
  et~al.}{2007}]{Kang2007}
{Kang} X.,  {van den Bosch} F.~C.,    {Pasquali} A.,  2007, \mnras, 381, 389

\bibitem[\protect\citeauthoryear{{Kauffmann}}{{Kauffmann}}{1996}]{Kauffmann199%
6}
{Kauffmann} G.,  1996, \mnras, 281, 487

\bibitem[\protect\citeauthoryear{{Kauffmann} \& {Charlot}}{{Kauffmann} \&
  {Charlot}}{1998}]{Kauffmann1998}
{Kauffmann} G.,  {Charlot} S.,  1998, \mnras, 297, L23

\bibitem[\protect\citeauthoryear{{Kennicutt}}{{Kennicutt}}{1983}]{Kennicutt198%
3}
{Kennicutt} R.~C.,  1983, \apj, 272, 54

\bibitem[\protect\citeauthoryear{{Khochfar} \& {Silk}}{{Khochfar} \&
  {Silk}}{2006}]{Khochfar2006}
{Khochfar} S.,  {Silk} J.,  2006, \mnras, 370, 902

\bibitem[\protect\citeauthoryear{{Kormendy}}{{Kormendy}}{1993}]{Kormendy1993}
{Kormendy} J.,  1993, in {Dejonghe} H.,  {Habing} H.~J.,  eds, {Galactic
  Bulges} Vol.~153 of IAU Symposium, {Kinematics of extragalactic bulges:
  evidence that some bulges are really disks}.
pp 209--+

\bibitem[\protect\citeauthoryear{{Kormendy} \& {Fisher}}{{Kormendy} \&
  {Fisher}}{2008}]{Kormendy2008}
{Kormendy} J.,  {Fisher} D.~B.,  2008, ArXiv e-prints

\bibitem[\protect\citeauthoryear{{Kormendy} \& {Gebhardt}}{{Kormendy} \&
  {Gebhardt}}{2001}]{Kormendy2001}
{Kormendy} J.,  {Gebhardt} K.,  2001, in {Wheeler} J.~C.,  {Martel} H.,  eds,
  20th Texas Symposium on relativistic astrophysics Vol.~586 of American
  Institute of Physics Conference Series, {Supermassive Black Holes in Galactic
  Nuclei (Plenary Talk)}.
pp 363--+

\bibitem[\protect\citeauthoryear{{Kormendy} \& {Illingworth}}{{Kormendy} \&
  {Illingworth}}{1982}]{Kormendy1982}
{Kormendy} J.,  {Illingworth} G.,  1982, \apj, 256, 460

\bibitem[\protect\citeauthoryear{{Kormendy} \& {Kennicutt} Jr}{{Kormendy} \&
  {Kennicutt}}{2004}]{Kormendy2004}
{Kormendy} J.,  {Kennicutt} Jr R.~C.,  2004, \araa, 42, 603

\bibitem[\protect\citeauthoryear{{Lacey}, {Baugh}, {Frenk}, {Silva}, {Granato}
  \& {Bressan}}{{Lacey} et~al.}{2008}]{Lacey08}
{Lacey} C.~G.,  {Baugh} C.~M.,  {Frenk} C.~S.,  {Silva} L.,  {Granato} G.~L.,
   {Bressan} A.,  2008, \mnras, 385, 1155

\bibitem[\protect\citeauthoryear{{Larson}, {Tinsley} \& {Caldwell}}{{Larson}
  et~al.}{1980}]{Larson1980}
{Larson} R.~B.,  {Tinsley} B.~M.,    {Caldwell} C.~N.,  1980, \apj, 237, 692

\bibitem[\protect\citeauthoryear{{Lemson} \& {Springel}}{{Lemson} \&
  {Springel}}{2006}]{Lemson2006b}
{Lemson} G.,  {Springel} V.,  2006, in {Gabriel} C.,  {Arviset} C.,  {Ponz} D.,
    {Enrique} S.,  eds, {Astronomical Data Analysis Software and Systems XV}
  Vol.~351 of {Astronomical Society of the Pacific Conference Series},
  {Cosmological Simulations in a Relational Database: Modelling and Storing
  Merger Trees}.
pp 212--+

\bibitem[\protect\citeauthoryear{{Lemson} \& {The Virgo Consortium}}{{Lemson}
  \& {The Virgo Consortium}}{2006}]{Lemson2006}
{Lemson} G.,  {The Virgo Consortium} 2006, ArXiv e-prints, astro-ph/0608019

{Magorrian} J. {\etal},  1998, \aj, 115, 2285

\bibitem[\protect\citeauthoryear{{Malbon}, {Baugh}, {Frenk} \&
  {Lacey}}{{Malbon} et~al.}{2007}]{Malbon2007}
{Malbon} R.~K.,  {Baugh} C.~M.,  {Frenk} C.~S.,    {Lacey} C.~G.,  2007,
  \mnras, 382, 1394

\bibitem[\protect\citeauthoryear{{Marconi} \& {Hunt}}{{Marconi} \&
  {Hunt}}{2003}]{Marconi2003}
{Marconi} A.,  {Hunt} L.~K.,  2003, \apj, 589, L21

\bibitem[\protect\citeauthoryear{{Marleau} \& Simard}{{Marleau} \&
  Simard}{1998}]{Marleau1998}
{Marleau} F.~R.,  Simard L.,  1998, \apj, 507, 585

\bibitem[\protect\citeauthoryear{{Mo}, {Mao} \& {White}}{{Mo}
  et~al.}{1998}]{Mo1998}
{Mo} H.~J.,  {Mao} S.,    {White} S. D.~M.,  1998, \mnras, 295, 319

\bibitem[\protect\citeauthoryear{{Moore}, {Katz}, {Lake}, {Dressler} \&
  {Oemler}~Jr}{{Moore} et~al.}{1996}]{Moore1996}
{Moore} B.,  {Katz} N.,  {Lake} G.,  {Dressler} A.,    {Oemler}~Jr A.,  1996,
  \nat, 379, 613

\bibitem[\protect\citeauthoryear{{Naab} \& {Burkert}}{{Naab} \&
  {Burkert}}{2003}]{Naab2003}
{Naab} T.,  {Burkert} A.,  2003, \apj, 597, 893

\bibitem[\protect\citeauthoryear{{Naab}, {Burkert} \& {Hernquist}}{{Naab}
  et~al.}{1999}]{Naab1999}
{Naab} T.,  {Burkert} A.,    {Hernquist} L.,  1999, \apj, 523, L133

\bibitem[\protect\citeauthoryear{{Naab}, {Jesseit} \& {Burkert}}{{Naab}
  et~al.}{2006}]{Naab2006}
{Naab} T.,  {Jesseit} R.,    {Burkert} A.,  2006, \mnras, 372, 839

\bibitem[\protect\citeauthoryear{{Nakamura}, {Fukugita}, {Yasuda}, {Loveday},
  {Brinkmann}, {Schneider}, {Shimasaku} \& {SubbaRao}}{{Nakamura}
  et~al.}{2003}]{Nakamura2003}
{Nakamura} O. {\etal},  2003, \apj, 125, 1682

\bibitem[\protect\citeauthoryear{{Negroponte} \& {White}}{{Negroponte} \&
  {White}}{1983}]{Negroponte83}
{Negroponte} J.,  {White} S.~D.~M.,  1983, \mnras, 205, 1009

\bibitem[\protect\citeauthoryear{{Papovich}, {Dickinson}, {Giavalisco},
  {Conselice} \& {Ferguson}}{{Papovich} et~al.}{2005}]{Papovich2005}
{Papovich} C.,  {Dickinson} M.,  {Giavalisco} M.,  {Conselice} C.~J.,
  {Ferguson} H.~C.,  2005, \apj, 631, 101

\bibitem[\protect\citeauthoryear{{Peebles}}{{Peebles}}{1969}]{Peebles1969}
{Peebles} P. J.~E.,  1969, \apj, 155, 393

\bibitem[\protect\citeauthoryear{{Postman}, {Franx}, {Cross}, {Holden}, {Ford}
  \& {Illingworth}}{{Postman} et~al.}{2005}]{Postman2005}
{Postman} M. {\etal},  2005, \apj, 623, 721

\bibitem[\protect\citeauthoryear{{Postman} \& {Geller}}{{Postman} \&
  {Geller}}{1984}]{Postman1984}
{Postman} M.,  {Geller} M.~J.,  1984, \apj, 281, 95

\bibitem[\protect\citeauthoryear{{Ravindranath}, {Ferguson}, {Conselice},
  {Giavalisco}, {Dickinson} \& {Chatzichristou}}{{Ravindranath}
  et~al.}{2004}]{Ravindranath2004}
{Ravindranath} S. {\etal},  2004 , \apj, 604, L9

\bibitem[\protect\citeauthoryear{{Sandage}}{{Sandage}}{1961}]{Sandage1961}
{Sandage} A.,  1961, {The Hubble atlas of galaxies}.
Washington: Carnegie Institution

\bibitem[\protect\citeauthoryear{{Smith}, {Loveday} \& {Cross}}{{Smith}
  et~al.}{2008}]{Smith2008}
{Smith} A.~J.,  {Loveday} J.,    {Cross} N.~J.~G.,  2008, ArXiv e-prints

\bibitem[\protect\citeauthoryear{{Somerville}, {Hopkins}, {Cox}, {Robertson} \&
  {Hernquist}}{{Somerville} et~al.}{2008}]{Somerville2008}
{Somerville} R.~S.,  {Hopkins} P.~F.,  {Cox} T.~J.,  {Robertson} B.~E.,
  {Hernquist} L.,  2008, ArXiv e-prints

\bibitem[\protect\citeauthoryear{{Spergel}, {Verde}, {Peiris}, {Komatsu},
  {Nolta} \& {Bennett}}{{Spergel} et~al.}{2003}]{Spergel2003}
{Spergel} D.~N. {\etal},  2003, \apjs, 148, 175

\bibitem[\protect\citeauthoryear{{Springel} \& {Hernquist}}{{Springel} \&
  {Hernquist}}{2005}]{Springel2005b}
{Springel} V.,  {Hernquist} L.,  2005, \apjl, 622, L9

\bibitem[\protect\citeauthoryear{{Springel}, {White}, {Jenkins}, {Frenk},
  {Yoshida} \& {Gao}}{{Springel} et~al.}{2005}]{Springel2005sup}
{Springel} V. {\etal},  2005, \nat, 435, 629

\bibitem[\protect\citeauthoryear{{Springel}, {White}, {Tormen} \&
  {Kaufmann}}{{Springel} et~al.}{2001}]{Springel2001}
{Springel} V.,  {White} S. D.~M.,  {Tormen} G.,    {Kaufmann} G.,  2001,
  \mnras, 328, 726

\bibitem[\protect\citeauthoryear{{Steinmetz} \& {Navarro}}{{Steinmetz} \&
  {Navarro}}{2002}]{Steinmetz2002}
{Steinmetz} M.,  {Navarro} J.,  2002, New.Astron., 7, 155

\bibitem[\protect\citeauthoryear{{Tasca} \& {White}}{{Tasca} \&
  {White}}{2005}]{Tasca2005}
{Tasca} L. A.~M.,  {White} S. D.~M.,  2005, \mnras, submitted
  (astro-ph/0507249)

\bibitem[\protect\citeauthoryear{{Toomre}}{{Toomre}}{1977}]{Toomre1977}
{Toomre} A.,  1977, in {Evolution of Galaxies and Stellar Populations} {Mergers
  and Some Consequences}.
pp 401--+

\bibitem[\protect\citeauthoryear{{Tran}, {Simard}, {Zabludoff} \&
  {Mulchaey}}{{Tran} et~al.}{2001}]{Tran2001}
{Tran} K.~V.,  {Simard} L.,  {Zabludoff} A.~I.,    {Mulchaey} J.~S.,  2001,
  \apj, 549, 172

\bibitem[\protect\citeauthoryear{{Van den Bergh}}{{Van den
  Bergh}}{1960}]{VDB1960}
{Van den Bergh} S.,  1960, \apj, 131, 215

\bibitem[\protect\citeauthoryear{{Van den Bergh}}{{Van den
  Bergh}}{2002}]{VanDenBergh2002}
{Van den Bergh} S.,  2002, \pasp, 114, 797

\bibitem[\protect\citeauthoryear{{White}}{{White}}{1978}]{White1978}
{White} S. D.~M.,  1978, \mnras, 184, 183

\bibitem[\protect\citeauthoryear{{White}}{{White}}{1984}]{White1984}
{White} S. D.~M.,  1984, \apj, 286, 38

\bibitem[\protect\citeauthoryear{{White} \& {Frenk}}{{White} \&
  {Frenk}}{1991}]{White1991}
{White} S. D.~M.,  {Frenk} C.~S.,  1991, \apj, 379, 52

\bibitem[\protect\citeauthoryear{{Zavala}, {Okamoto} \& {Frenk}}{{Zavala}
  et~al.}{2008}]{Zavala2008}
{Zavala} J.,  {Okamoto} T.,    {Frenk} C.~S.,  2008, \mnras, 387, 364
\end{thebibliography}

\label{lastpage}
\end{document}